\documentclass[a4paper,twocolumn,english,pra,eqsecnum,showpacs]{revtex4}
\usepackage[T1]{fontenc}
\usepackage[latin1]{inputenc}
\usepackage{graphicx}

\makeatletter

\providecommand{\tabularnewline}{\\}

\usepackage{ae,aecompl}

\usepackage{bm}

\usepackage{babel}
\makeatother
\begin{document}

\title{Gaussian quantum operator representation for bosons }

\author{Joel F. Corney and Peter D. Drummond}

\affiliation{ARC Centre for Quantum Atom Optics, University of Queensland, St
Lucia 4072, Queensland, Australia.}

\begin{abstract}
We introduce a Gaussian quantum operator representation, using the
most general possible multi-mode Gaussian operator basis. The representation
unifies and substantially extends existing phase-space representations
of density matrices for Bose systems, and also includes generalized
squeezed-state and thermal bases. It enables first-principles dynamical
or equilibrium calculations in quantum many-body systems, with quantum
uncertainties appearing as dynamical objects. Any quadratic Liouville
equation for the density operator results in a purely deterministic
time evolution. Any cubic or quartic master equation can be treated
using stochastic methods.
\end{abstract}


\maketitle

\section{Introduction}

In this paper we develop a general multi-mode Gaussian representation
for a density matrix of bosons. As well as classical phase-space variables
like $(\mathbf{x},\mathbf{p})$, the representation utilizes a dynamical
space of quantum uncertainties or co-variances. The extended phase
space accommodates more efficiently the content of a quantum state,
and allows the physics of many kinds of problems to be incorporated
into the basis itself. The Gaussian expansion technique unifies and
greatly extends all the previous Gaussian-like phase-space representations
used for bosons, including the Wigner, $Q$, $P$, positive-$P,$
and squeezed-state expansions. The operator basis also includes non-Hermitian
Gaussian operators, which are not density matrices themselves, but
can form part of a probabilistic expansion of a physical density matrix.
Unlike previous approaches, \emph{any} initial state is found to evolve
with a deterministic time-evolution under \emph{any} quadratic Hamiltonian
or master equation. 

The complexity of many-body quantum physics is manifest in the enormity
of the Hilbert space of systems with even modest numbers of particles.
This complexity makes it prohibitively difficult to simulate quantum
dynamics with exact precision: no digital computer is large enough
to store the dynamically evolving state. However, if a finite precision
is permitted, then quantum dynamical calculations are possible, through
what are known as phase-space methods. These methods represent the
evolving quantum state as probability distributions on some suitable
phase space, which can be sampled via stochastic techniques. The mapping
to phase space can be made to be exact. Thus the precision of the
final result is limited only by sampling error, which can usually
be reliably estimated and which can be reduced by an increased number
of stochastic paths.

Arbitrary quantum mechanical evolution cannot be represented (even
probabilistically) on a phase space as is usually defined. Thus the
extended phase spaces employed here are a generalization of conventional
phase spaces in several ways. First, it is a \emph{quantum} phase
space, in which points can correspond to states with intrinsic uncertainty.
Heisenberg's uncertainty relations can thus be satisfied in this way,
or more generally by considering genuine probability distributions
over phase space, to be sampled stochastically. Second, the phase
space is of \emph{double dimension}, where classically real variables,
such as $\mathbf{x}$ and $\mathbf{p}$, now range over the complex
plane. This allows arbitrary quantum \emph{evolution} to be sampled
stochastically. Third, stochastic \emph{gauge} variables are included.
These arbitrary quantities do not affect the physical results, but
they can be used to overcome problems in the stochastic sampling.
Fourth, the phase space includes the set of second-order moments,
or \emph{covariances}. A phase space that is enlarged in this way
is able to accommodate more information about a general quantum state
in a single point. In particular, any state (pure or mixed) with Gaussian
statistics can be represented as a single point in this phase space.

The Gaussian representation provides a link between phase-space methods
and approximate methods used in many-body theory, which frequently
treat normal and anomalous correlations or Green's functions as dynamical
objects\cite{BaymKad}. As well as being applicable to quantum optics
and quantum information, a  strong motivation for this representation
is the striking experimental observation of BEC in ultra-cold atomic
systems\cite{obsex1}. Already the term `atom laser' is widely used,
and experimental observation of quantum statistics in these systems
is underway. Yet there is a problem in using previous quantum optics
formalisms to calculate coherence properties in atom optics: interactions
are generally much stronger with atoms than they are with photons,
relative to the damping rate. The consequence of this is that one
must anticipate larger departures from 'semiclassical', coherent-state
behaviour in atomic systems. 

The present paper includes these nonclassical and incoherent effects
at the level of the basis for the operator representation itself.
The purpose of employing a Gaussian basis set is not only to enlarge
the parameter set (to hold more information about the quantum state),
but also to include in the basis states that are a close match to
the actual states that are likely to occur in interesting systems,
such as dilute gases. The pay-off for increasing the parameter set
is more efficient sampling of the dynamically evolving or equilibrium
states of many-body systems.

The idea of coherent states as a quasi-classical basis for quantum
mechanics originated with Schr\"odinger\cite{Scrod}. Subsequently,
Wigner \cite{Wig-Wigner} introduced a distribution for quantum density
matrices. This method was a phase-space mapping with classical dimensions
and employed a symmetrically ordered operator correspondence principle.
Later developments included the antinormally ordered $Q$ distribution\cite{Hus-Q},
a normally ordered expansion called the diagonal $P$ distribution\cite{Gla-P},
methods that interpolate between these classical phase-space distributions\cite{CG-Q,Agarwal},
and diagonal squeezed-state representations\cite{SchCav84}. Each
of these expansions either employs an explicit Gaussian density matrix
basis or is related to one that does by convolution. They are suitable
for phase-space representations of quantum states because of the overcompleteness
of the set of coherent states on which they are based. 

Arbitrary pure states of bosons with a Gaussian wave-function or Wigner
representation are often called the squeezed states\cite{Squeeze}.
These are a superset of the coherent states, and were investigated
by Bogoliubov\cite{Bogol} to approximately represent the ground state
of an interacting Bose-Einstein condensate - as well as in much recent
work in quantum optics\cite{QOSqueeze}. Diagonal expansions analogous
to the diagonal $P$-representation have been introduced using a basis
of squeezed-state projectors, typically with a fixed squeezing parameter\cite{SchCav84}.
However, these have not generally resulted in useful dynamical applications,
as they do not overcome the problems inherent in using a diagonal
basis, as we discuss below. 

In operator representations, one must utilize a complete basis in
the Hilbert space of density operators, rather than in the Hilbert
space of pure states. Thermal density matrices, for example, are not
pure states, but do have a Gaussian $P$ representation and Wigner
function. To include all three types of commonly used Gaussian states
- the coherent, squeezed and thermal states - one can define a Gaussian
\emph{state} as a density matrix having a Gaussian positive-$P$ or
Wigner representation\cite{Lindbladgauss}. This definition also includes
displaced and squeezed thermal states. Gaussian states have been investigated
extensively in quantum information and quantum entanglement\cite{GaussQI}.
It has been shown that an initial Gaussian state will remain Gaussian
under linear evolution\cite{Bartlett}. 

However, the Gaussian density matrices that correspond to physical
states do \emph{not} by themselves form a complete basis for the time-evolution
of all quantum density matrices. This problem, inherent in all diagonal
expansions, is related to known issues in constructing quantum--classical
correspondences\cite{Neumann}, and is caused by the non-positive-definite
nature of the local propagator in a classical phase space. It is manifest
in the fact that there is generally no equivalent Fokker-Planck equation
(with a positive-definite diffusion matrix) that generates the quantum
time-evolution, and hence no corresponding stochastic differential
equation that can be efficiently simulated numerically. This difficulty
occurs in nearly all cases except free fields, and represents a substantial
limitation in the use of these diagonal expansion methods for exact
simulation of the quantum dynamics of interacting systems.

These problems can be solved by use of nonclassical phase spaces,
which correspond to expansions in non-Hermitian bases of operators
\emph{}(rather than just physical density matrices). One established
example is the non-diagonal positive-$P$\cite{DG-PosP} representation.
The non-Hermitian basis in this case generates a representation with
a positive propagator, which allows the use of stochastic methods
to sample the quantum dynamics. By extending the expansion to include
a stochastic gauge freedom in these evolution equations, one can select
the most compact possible time-evolution equation\cite{gauge_paper,Paris1,Plimak}.
With an appropriate gauge choice, this method is exact for a large
class of nonlinear Hamiltonians, since it eliminates boundary terms
that can otherwise arise\cite{GGD-Validity}. The general Gaussian
representation used here also includes these features, and extends
them to allow treatment of any Hamiltonian or master equation with
up to fourth-order polynomial terms.

Other methods of theoretical physics that have comparable goals are
the path-integral techniques of quantum field theory\cite{wilson:74,Ceperley},
and density functional methods\cite{Kohn} that are widely used to
treat atomic and molecular systems. The first of these is exact in
principle, but is almost exclusively used in imaginary time calculations
of canonical ensembles or ground states due to the notorious phase
problem. The second method has similarities with our approach in that
it also utilizes a density as we do. However, density functionals
are normally combined with approximations like the local density approximation.
Gaussian representation methods have the advantage that they can treat
both real and imaginary time evolution. In addition, the technique
is exact in principle, provided boundary terms vanish on partial integration.

In section \ref{sec:Gaussian--expansions}, we define general Gaussian
operators for a density operator expansion and introduce a compact
notation for these operators, either in terms of mode operators or
quantum fields. In section \ref{sec:Gaussian-expectation-values},
we calculate the moments of the general Gaussian representation, relating
them to physical quantities as well as to the moments of previous
representations. Section \ref{sec:Gaussian-differential-identities}
gives the necessary identities that enable first-principles quantum
calculations with these representations. 

Equation (\ref{eq:Matrixidentities}) summarizes the relevant operator
mappings, and constitutes a key result of the paper. 

We give a number of examples in section \ref{sec:Examples} of specific
pure and mixed states (and their non-Hermitian generalizations) that
are included in the basis, and we give simplified versions of important
identities for these cases. Section \ref{sec:Time-evolution} describes
how the Gaussian representation can be used to deal with evolution
in either real or imaginary time. In particular, we show how it can
be used to solve exactly any master equation that is quadratic in
annihilation and creation operators. Some useful normalization integrals
and reordering identities for the Gaussian operators are proved in
the Appendices.

In a subsequent paper, we will apply these methods to systems with
nonlinear evolution.

\section{The Gaussian representation}

\label{sec:Gaussian--expansions}
\begin{figure}
\includegraphics[%
  scale=0.5]{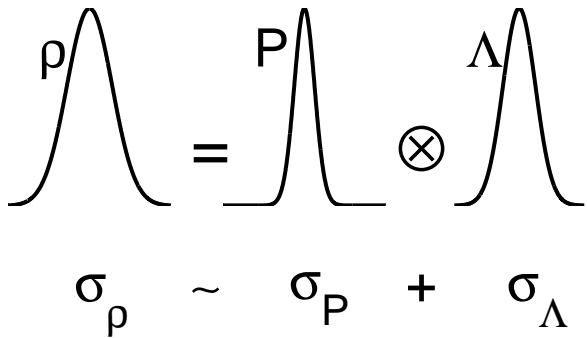}

\caption{The density-operator expansion in Eq. (\ref{eq:general-expansion})
can be interpreted as a convolution of the probability distribution
$P$ with the underlying distribution of the basis. The uncertainty
or spread of the physical state, indicated by the variance $\sigma_{\rho}$,
is shared between the distribution variance $\sigma_{P}$ and the
basis variance $\sigma_{\Lambda}$.}

\label{figure1}
\end{figure}

The representations that give exact
mappings between operator equations and stochastic equations -- an
essential step toward representing operator dynamics in large Hilbert
spaces -- are stochastic gauge expansions\cite{gauge_paper,Paris1,Plimak}
on a nonclassical phase space. Here, the generic expansion is written
down in terms of a complete set of operators that are typically non-Hermitian.
This leads to the typical form:

\begin{equation}
\widehat{\rho}(t)=\int P(\overrightarrow{\lambda},t)\widehat{\Lambda}(\overrightarrow{\lambda})d\overrightarrow{\lambda}\,\,,\label{eq:general-expansion}\end{equation}
where $P(\overrightarrow{\lambda},t)$ is a probability distribution,
$\widehat{\Lambda}$ is a suitable basis for the class of density
matrices being considered, and $d\overrightarrow{\lambda}$ is the
integration measure for the corresponding generalized phase-space
coordinate $\overrightarrow{\lambda}$. See Fig.~\ref{figure1} for
a conceptual illustration of this expansion. 

In phase-space methods, it is the distribution $P$ that is sampled
stochastically. Therefore if the basis resembles the typical physical
states of a system, the sampling error will be minimized, and if the
state coincides exactly with an element of the basis, then the distribution
will be a delta function, with consequently no sampling error. A Wigner
or $Q$-function basis, for example, generates a broad distribution
even for minimum uncertainty states. A general Gaussian basis, on
the other hand, can generate a delta-function distribution not only
for any minimum uncertainty state, but also for the ground states
of non-interacting finite-temperature systems.

\subsection{Gaussian operator basis}

In this paper, we define the operator basis $\widehat{\Lambda}$ to
be the most general Gaussian operator basis. The motivation for using
the most general possible basis set is that when the basis set members
nearly match the states of interest, the resulting distributions are
more compact and have lower sampling errors in a Monte-Carlo or stochastic
calculation. In addition, a larger basis allows more choice of mappings,
so that lower-order differential correspondences can be utilized.
In some cases, a large basis set can increase computational memory
requirements, as more parameters are needed. This disadvantage is
outweighed when there is a substantial decrease in sampling error,
due to the use of a more physically appropriate basis. By choosing
a general Gaussian operator basis, rather than just a basis of Gaussian
density matrices, one has the additional advantage of a complete representation
for all non-Gaussian density matrices as well. 

If $\widehat{\bm a}$ is a column vector of $M$ bosonic annihilation
operators, and $\widehat{\bm a}^{\dagger}$ the corresponding row
vector of creation operators, their commutation relations are:\begin{equation}
\left[\widehat{a}_{k},\widehat{a}_{j}^{\dagger}\right]=\delta_{kj}\,\,.\label{eq:a-com}\end{equation}
We define a Gaussian operator as an exponential of an arbitrary quadratic
form in annihilation and creation operators (or, equivalently, a quadratic
form in position and momentum operators). 

The simplest way to achieve this is to introduce extended $2M$-vectors
of c-numbers and operators: $\underline{\alpha}=(\,\bm\alpha,(\bm\alpha^{+})^{T})$,
and $\underline{\widehat{a}}=(\widehat{\bm a},(\widehat{\bm a}^{\dagger})^{T})$,
with adjoints defined as $\underline{\alpha}^{+}=(\bm\alpha^{+},\bm\alpha^{T})$
and $\underline{\widehat{a}}^{\dagger}=(\widehat{\bm a}^{\dagger},\widehat{\bm a}^{T})$,
together with a relative operator displacement of \begin{equation}
\delta\underline{\widehat{a}}=\underline{\widehat{a}}-\underline{\alpha}=\left(\begin{array}{c}
\widehat{a}_{1}\\
\vdots\\
\widehat{a}_{M}\\
\widehat{a}_{1}^{\dagger}\\
\vdots\\
\widehat{a}_{M}^{\dagger}\end{array}\right)-\left(\begin{array}{c}
\alpha_{1}\\
\vdots\\
\alpha_{M}\\
\alpha_{1}^{+}\\
\vdots\\
\alpha_{M}^{+}\end{array}\right)\,\,.\label{eq:delta-a}\end{equation}
 These extended vectors are indexed where necessary with Greek indices:
$\mu=1,...,2M$. 

A general Gaussian operator is now an exponential of a general quadratic
form in the $2M$-vector mode operator $\delta\underline{\widehat{a}}$.
For algebraic reasons, it is useful to employ normal ordering, and
to introduce a compact notation using a generalized covariance $\underline{\underline{\sigma}}$: 

\fbox{\parbox{0.9\columnwidth}{

\begin{eqnarray}
\widehat{\Lambda}(\overrightarrow{\lambda}) & = & \frac{\Omega}{\sqrt{\left|\underline{\underline{\sigma}}\right|}}:\exp\left[-\delta\underline{\widehat{a}}^{\dagger}\underline{\underline{\sigma}}^{-1}\delta\underline{\widehat{a}}/2\right]:\,\,.\label{eq:Gaussbasis}\end{eqnarray}
}}

~

Here the normalization factor involving $\sqrt{\left|\underline{\underline{\sigma}}\right|}$
is introduced to simplify identities that occur later, and plays a
very similar role to the exactly analogous normalization factor that
occurs in the classical Gaussian distribution of probability theory.
The $2M\times2M$ covariance matrix is conveniently parameterized
in terms of $M\times M$ submatrices as: \begin{equation}
\underline{\underline{\sigma}}=\left[\begin{array}{cc}
\mathbf{I}+\mathbf{n} & \mathbf{m}\\
\mathbf{m}^{+} & \mathbf{I}+\mathbf{n}^{T}\end{array}\right]\,\,,\label{Covariance}\end{equation}
where $\mathbf{n}$ is a complex $M\times M$ matrix and $\mathbf{m},\mathbf{m}^{+}$
are two independent symmetric complex $M\times M$ matrices. 

With this choice, the covariance has a type of generalized Hermitian
symmetry in which $\sigma_{\mu\nu}=\sigma_{\nu+M,\mu+M}$, provided
we interpret the matrix indices as cyclic in the sense that $\nu\sim\nu+2M$.
This can also be written as $\underline{\underline{\sigma}}=\underline{\underline{\sigma}}^{+},$with
the definition that:

\begin{equation}
\left[\begin{array}{cc}
\mathbf{a} & \mathbf{b}\\
\mathbf{c} & \mathbf{d}\end{array}\right]^{+}\equiv\left[\begin{array}{cc}
\mathbf{d} & \mathbf{c}\\
\mathbf{b} & \mathbf{a}\end{array}\right]^{T}\,\,.\label{eq:sym}\end{equation}
This definition implies that we intend the `$^{+}$' superscript to
define an operation on the covariance matrix which is equivalent to
Hermitian conjugation of the underlying operators. If the Gaussian
operator is in fact an Hermitian operator, then so is the corresponding
covariance matrix. In this case, the `$^{+}$' superscript is identical
to ordinary Hermitian conjugation. The generalized Hermitian symmetry
of the covariance means that all elements of the number correlation
$\mathbf{n}$ appear twice, as do all except the diagonal elements
of the squeezing correlations $\mathbf{m},\mathbf{m}^{+}$. 

The use of normal ordering allows simple operator identities to be
obtained, but can easily be related to more commonly used unordered
parameterizations. The Gaussian operators include as special cases
the density matrices of many useful and well-known physical states.
For example, they include the thermal states of a Bose-Einstein distribution,
the coherent states, and the squeezed states. They also include many
more states than these, like the off-diagonal coherent state projectors
used in the positive-$P$ expansion, which are not density operators
themselves, but can be used to expand density operators. The details
are given in section \ref{sec:Examples}.

\subsection{Extended phase space}

\label{sub:Extended-phase-space}

The representation phase space is thus extended to \begin{equation}
\overrightarrow{\lambda}=(\Omega,\bm\alpha,\bm\alpha^{+},\mathbf{n},\mathbf{m},\mathbf{m}^{+})\,\,.\label{eq:lambda}\end{equation}
The complex amplitude $\Omega$, which appears in the normalisation,
acts as a dynamical weight on different stochastic trajectories. It
is useful in calculations in which the normalisation of the density
matrix is not intrinsically preserved, such as canonical ensemble
calculations, and also enables stochastic gauges to be included. 

The complex vectors $\bm\alpha$ and $\bm\alpha^{+}$ give the generalized
coherent amplitudes for each mode: $\bm\alpha$ defines the amplitudes
of annihilation operators $\widehat{\bm a}$ , while its `conjugate'
$\bm\alpha^{+}$ defines the amplitudes of the creation operators
$\widehat{\bm a}^{\dagger}$. The matrix $\mathbf{n}$ gives the number,
or normal, correlations between each pair of modes. The squeezing,
or anomalous, correlations between each pair of modes are given by
$\mathbf{m}$ and $\mathbf{m}^{+}$: the matrix $\mathbf{m}$ defines
the correlations of annihilation operators, while its `conjugate'
$\mathbf{m}^{+}$ defines the correlations of the creation operators.
These physical interpretations of the phase-space variables are supported
by the results of Section \ref{sec:Gaussian-expectation-values},
where we rigorously establish the connection of the phase-space variables
to physical quantities.

In general, apart from the complex amplitude $\Omega$, the total
number of complex parameters needed to specify the normalized $M$-mode
Gaussian operator is:

\begin{equation}
p=M(2+3M)\,\,.\label{eq:param}\end{equation}
 Hence the phase-space variables can be written as $\overrightarrow{\lambda}=(\lambda_{0},\lambda_{1},...,\lambda_{p})$,
with the corresponding integration measure as $d\overrightarrow{\lambda}=d^{2(p+1)}\overrightarrow{\lambda}$.

\subsection{Gaussian field operators }

\label{sub:Gaussian-field-operators}

The above results define a completely general Gaussian operator in
terms of arbitrary bosonic annihilation and creation operators, without
reference to the field involved. It is sometimes useful to compare
this to a field theoretic notation, in which we explicitly use a coordinate-space
integral to define the correlations. This provides a means to extend
operator representation theory for fields\cite{Carter-Drummond,Kennedy,becpf,Drummond-Corney}
to more general basis sets. In a quantization volume $V$, one can
expand:

\begin{eqnarray}
\widehat{\Psi}_{j}(\mathbf{x}) & = & \frac{1}{\sqrt{V}}\sum_{\mathbf{k}}\widehat{a}_{\mathbf{k},j}e^{i\mathbf{k}.\mathbf{x}}\nonumber \\
\widehat{\Psi}_{j}^{\dagger}(\mathbf{x}) & = & \frac{1}{\sqrt{V}}\sum_{\mathbf{k}}\widehat{a}_{\mathbf{k},j}^{\dagger}e^{-i\mathbf{k}.\mathbf{x}}\,,\label{eq:field}\end{eqnarray}
where the field commutators are\begin{equation}
\left[\widehat{\Psi}_{j}(\mathbf{x}),\widehat{\Psi}_{j'}^{\dagger}(\mathbf{x}')\right]=\delta_{jj'}\delta(\mathbf{x}-\mathbf{x}')\,\,.\label{eq:field-com}\end{equation}

With this notation, the quadratic term in the Gaussian exponent becomes:\begin{equation}
\delta\underline{\widehat{a}}^{\dagger}\underline{\underline{\sigma}}^{-1}\delta\underline{\widehat{a}}=\int\int\delta\underline{\widehat{\Psi}}^{\dagger}(\mathbf{x})\underline{\underline{\sigma}}^{-1}(\mathbf{x},\mathbf{y})\delta\underline{\widehat{\Psi}}(\mathbf{y})d^{3}\mathbf{x}d^{3}\mathbf{y}\,\,,\label{eq:field-quad}\end{equation}
where we have introduced the extended vector $\underline{\widehat{\Psi}}(\mathbf{x})=(\widehat{\bm\Psi},\left(\widehat{\bm\Psi}^{\dagger}\right)^{T})$,
and $\delta\underline{\widehat{\Psi}}(\mathbf{x})=\underline{\widehat{\Psi}}(\mathbf{x})-\underline{\Psi}(\mathbf{x})$
which is the operator fluctuation relative to the coherent displacement
or classical mean field. If we index the extended vector as $\Psi_{js}$
, where $s=-1(1)$ for the first and second parts respectively, this
Fourier transform can be written compactly as: \begin{equation}
\Psi_{js}(\mathbf{x})=\frac{1}{\sqrt{V}}\sum_{\mathbf{k}}\alpha_{\mathbf{k}js}e^{-is\mathbf{k}.\mathbf{x}}\,\,.\label{eq:FT-combine}\end{equation}
The notation $\underline{\underline{\sigma}}^{-1}(\mathbf{x},\mathbf{y})$
indicates a functional matrix inverse where:

\begin{equation}
\int\underline{\underline{\sigma}}^{-1}(\mathbf{x},\mathbf{y})\underline{\underline{\sigma}}(\mathbf{y},\mathbf{x}')d^{3}\mathbf{y}=\underline{\underline{I}}\delta(\mathbf{x}-\mathbf{x}')\,\,,\label{eq:inverse}\end{equation}
and the relationship to the previous cross-variance matrix is that:\begin{equation}
\sigma_{js,j's'}(\mathbf{x},\mathbf{y})=\frac{1}{V}\sum_{\mathbf{k}}\sum_{\mathbf{k}'}\sigma_{\mathbf{k}js,\mathbf{k}'j's'}e^{-i(s\mathbf{k}.\mathbf{x}-s'\mathbf{k}'.\mathbf{x}')}\,.\label{eq:relation}\end{equation}

In the standard terminology of many-body theory and field theory\cite{BaymKad},
these field variances are generalized equal-time Green's functions,
and can be written as:\begin{equation}
\underline{\underline{\sigma}}(\mathbf{x},\mathbf{x}')=\left[\begin{array}{cc}
\mathbf{I}\delta(\mathbf{x},\mathbf{x}')+\mathbf{n}(\mathbf{x},\mathbf{x}')\,\,\, & \mathbf{m}(\mathbf{x},\mathbf{x}')\\
\mathbf{m}(\mathbf{x},\mathbf{x}')^{+} & \mathbf{I}\delta(\mathbf{x},\mathbf{x}')+\mathbf{n}^{T}(\mathbf{x}',\mathbf{x})\end{array}\right]\,\,.\label{fieldCovariance}\end{equation}

We shall show in the next section that these indeed correspond to
field correlation functions in the case that the field state is able
to be represented as a single Gaussian. More generally, one must consider
a probability distribution over different coherent fields and Green's
functions or variances, in order to construct the overall density
matrix.

\section{Gaussian expectation values}

\label{sec:Gaussian-expectation-values}In order to use the Gaussian
operator basis, a number of basic identities are needed. In this section,
we derive relations between operator expectation values and moments
of the distribution. Such moments also show how the general Gaussian
representation incorporates the previously used methods.

\subsection{Gaussian trace }

The trace of a generalized Gaussian is needed to normalize the density
matrix. The trace is most readily calculated by using a well-known
coherent-state identity\cite{Gla-P}, 

\begin{equation}
{\rm Tr}\left[\widehat{\Lambda}\right]=\int\left\langle \mathbf{z}\right|\widehat{\Lambda}\left|\mathbf{z}\right\rangle \frac{d^{2M}\mathbf{z}}{\pi^{M}}\,.\label{eq:coherent-trace}\end{equation}
Here we define $\mathbf{z}=(z_{1},...,z_{M})$. Next, introducing
extended vectors $\underline{z}=(\mathbf{z},\mathbf{z}^{\ast})^{T}$,
$\underline{z}^{+}=(\mathbf{z}^{*},\mathbf{z})$, $\delta\underline{z}=\underline{z}-\underline{\alpha}$,
and using the eigenvalue property of coherent states, $\widehat{\bm a}\left|\mathbf{z}\right\rangle =\mathbf{z}\left|\mathbf{z}\right\rangle $,
we find that:

\begin{equation}
{\rm Tr}\left[\widehat{\Lambda}\right]=\frac{\Omega\int d^{2M}\mathbf{z}\exp\left[-\delta\underline{z}^{+}\underline{\underline{\sigma}}^{-1}\delta\underline{z}/2\right]}{\pi^{M}\sqrt{\left|\underline{\underline{\sigma}}\right|}}\,\,.\label{Trace}\end{equation}
The normalizing factor can now be recognized as the determinant expression
arising in a classical Gaussian. For example, in the single-mode case,
one obtains for the normalizing determinant that:

\begin{equation}
\frac{1}{\sqrt{\left|\underline{\underline{\sigma}}\right|}}=\frac{1}{\sqrt{(1+n)^{2}-mm^{+}}\,}\,.\label{singlemodenorm}\end{equation}
We can thus calculate the value of the normalization from standard
Gaussian integrals, as detailed in Appendix \ref{sec:Gaussian-Integrals},
provided $\underline{\underline{\sigma}}$ has eigenvalues with a
positive real part. The result is:\begin{equation}
{\rm Tr}\left[\widehat{\Lambda}\right]=\Omega\,\,.\label{eq:trace-norm}\end{equation}

Thus for $\Lambda$ itself to correspond to a normalised density matrix,
we must have $\Omega=1$. In a general expansion of a density matrix,
there may be terms which do not have this normalisation; with the
proviso the average weight still be $\left\langle \Omega\right\rangle =1$.
This freedom of having different weights on different members of the
ensemble provides a way of introducing gauge variables, which can
be used to improve the efficiency of the stochastic sampling but which
do not affect of the average result. The weight also allows calculations
to be performed in which the trace of the density matrix is not preserved,
as in canonical-ensemble calculations.

\subsection{Expectation values}

\label{sub:Expectation-values}

Given a density matrix expanded in Gaussian operators, it is essential
to be able to calculate operator expectation values. This can be achieved
most readily if the operator $\widehat{O}$ is written in antinormally
ordered form, as:\begin{equation}
\widehat{O}=\sum_{i}\widehat{u}_{i}(\widehat{\bm a})\widehat{v}_{i}(\widehat{\bm a}^{\dagger})=o(\underline{\widehat{a}})\,\,.\label{eq:anti-norm}\end{equation}

Since the density matrix expansion is normally ordered by definition,
the cyclic properties of a trace allows the expectation value of any
antinormally ordered operator to be re-arranged as a completely normally
ordered form. Hence, following a similar coherent-state expansion
procedure to that the previous subsection, we arrive at an expression
analogous to the kernel trace, Eq (\ref{Trace}): 

\begin{eqnarray}
\left\langle \widehat{O}\right\rangle  & = & \frac{{\rm Tr}\left[\sum_{i}\widehat{v}_{i}(\widehat{\bm a}^{\dagger})\widehat{\rho}\widehat{u}_{i}(\widehat{\bm a})\right]}{{\rm Tr}\left[\widehat{\rho}\right]}\nonumber \\
 & = & \frac{\int P(\overrightarrow{\lambda})O(\overrightarrow{\lambda})\Omega d\overrightarrow{\lambda}}{\int P(\overrightarrow{\lambda})\Omega d\overrightarrow{\lambda}}\,\,,\nonumber \\
 & = & \left\langle O(\overrightarrow{\lambda})\right\rangle _{P}\,\,.\end{eqnarray}
Here we have introduced an equivalence between the quantum expectation
value $\left\langle \widehat{O}\right\rangle $, and the weighted
probabilistic average $\left\langle O(\overrightarrow{\lambda})\right\rangle _{P}$.
This is an \emph{antinormally} ordered c-number operator equivalence
in phase space of $O(\overrightarrow{\lambda})\sim\widehat{O}$, where
the eigenvalue relations of coherent states are utilized to obtain:\begin{eqnarray}
O(\overrightarrow{\lambda}) & = & \frac{\int d^{2M}\mathbf{z}o(\underline{z})\exp\left[-\delta\underline{z}^{+}\underline{\underline{\sigma}}^{-1}\delta\underline{z}/2\right]}{\pi^{M}\sqrt{\left|\underline{\underline{\sigma}}\right|}}\,\,\nonumber \\
 & = & \left\langle o(\underline{z})\right\rangle _{\overrightarrow{\lambda}}\,\,.\label{gaussian-moment}\end{eqnarray}

Here $\left\langle o(\underline{z})\right\rangle _{\overrightarrow{\lambda}}$represents
the \emph{classical} Gaussian average of the c-number function $o(\underline{z})$.
In other words, all quantum averages are now obtained by a convolution
of a classical Gaussian average with a width $\sigma_{\Lambda}$ that
depends on the kernel parameter $\overrightarrow{\lambda}$, together
with a probabilistic average over $\overrightarrow{\lambda}$, with
a width $\sigma_{P}$ that depends on the phase-space distribution
$P(\overrightarrow{\lambda})$. The situation is depicted schematically
in Fig \ref{figure1}.

Consider the first-order moment where $\widehat{O}=\widehat{a}_{\mu}$.
This is straightforward, as $o(\underline{z})=z_{\mu}$, and the Gaussian
average of $o(\underline{z})$ is simply the Gaussian mean $\alpha_{\mu}$:\begin{equation}
\left\langle \widehat{a}_{\mu}\right\rangle =\overline{a}_{\mu}=\left\langle \alpha_{\mu}\right\rangle _{P}\,\,.\label{eq:linear}\end{equation}

More generally, to calculate the antinormally ordered moment $o(\underline{\widehat{a}})=\{\widehat{a}_{\mu_{1}}\widehat{a}_{\mu_{2}}....\widehat{a}_{\mu_{n}}\}$,
one must calculate the corresponding Gaussian moment $o(\underline{z})=z_{\mu_{1}}z_{\mu_{2}}...z_{\mu_{n}}$.
This is most easily achieved by use of the moment-generating function
for the Gaussian distribution in Eq. (\ref{gaussian-moment}), which
is\begin{eqnarray}
\chi_{A}(\underline{t},\overrightarrow{\lambda}) & = & e^{\underline{t}^{\ast}\underline{\alpha}+\underline{t^{\ast}}\,\underline{\underline{\sigma}}\,\underline{t}/2},\end{eqnarray}
where $\underline{t}=(t_{1},...,t_{M},t_{1}^{\ast},...,t_{M}^{\ast})=({\bf t},{\bf t}^{\ast T})$.
General moments of the Gaussian distribution are then given by:\begin{eqnarray}
\left\langle o(\underline{z})\right\rangle _{\overrightarrow{\lambda}} & = & \left.\frac{\partial^{n}}{\partial t_{\mu_{1}}^{\ast}\partial t_{\mu_{2}}^{\ast}...\partial t_{\mu_{n}}^{\ast}}\chi_{A}(\underline{t},\overrightarrow{\lambda})\right|_{\underline{t}=\underline{0}},\end{eqnarray}
where it must be remembered that the adjoint vector $\underline{t}^{\ast}$
is not independent of $\underline{t}$. We note that averaging the
moment-generating function over the distribution $P(\overrightarrow{\lambda})$
gives the antinormal quantum characteristic function of the density
operator:\begin{eqnarray}
\chi_{A}({\bf t},{\bf t}^{\ast}) & \equiv & {\rm Tr}\left\{ \widehat{\rho}e^{{\bf t}^{\ast}\,\widehat{{\bf a}}}\, e^{\widehat{{\bf a}}^{\dagger}\,{\bf t}}\right\} \nonumber \\
 & = & \int P(\overrightarrow{\lambda})\Omega e^{\underline{t}^{\ast}\underline{\alpha}+\underline{t^{\ast}}\,\underline{\underline{\sigma}}\,\underline{t}/2}d\overrightarrow{\lambda}\,\,.\end{eqnarray}
 This equation is an alternative way of (implicitly) defining the
Gaussian $P$ distribution as a function whose generalized Fourier
transform is equal to the quantum characteristic function.

As an example of a moment calculation, one obtains the c-number operator
equivalence for general \emph{normally ordered} quadratic term as:\begin{equation}
\left\langle :\widehat{a}_{\mu}\widehat{a}_{\nu}^{\dagger}:\right\rangle =\left\langle \alpha_{\mu}\alpha_{\nu}^{+}+\sigma_{\mu\nu}^{N}\right\rangle _{P}\,,\label{eq:variance}\end{equation}
where we have introduced the normally ordered covariance:

\begin{equation}
\underline{\underline{\sigma}}^{N}=\underline{\underline{\sigma}}-\underline{\underline{I}}\,.\label{eq:normalcov}\end{equation}

Writing these out in more detail, we obtain the following central
results for calculating normally ordered observables up to quadratic
order:\\

\fbox{\parbox{0.8\columnwidth}{\begin{eqnarray}
\left\langle \widehat{a}_{i}\right\rangle  & = & \left\langle \alpha_{i}\right\rangle _{P}\nonumber \\
\left\langle \widehat{a}_{i}^{\dagger}\right\rangle  & = & \left\langle \alpha_{i}^{+}\right\rangle _{P}\,\nonumber \\
\left\langle \widehat{a}_{i}\widehat{a}_{j}\right\rangle  & = & \left\langle \alpha_{i}\alpha_{j}+m_{ij}\right\rangle _{P}\nonumber \\
\left\langle :\widehat{a}_{i}\widehat{a}_{j}^{\dagger}:\right\rangle  & = & \left\langle \alpha_{i}\alpha_{j}^{+}+n_{ij}\right\rangle _{P}\,\nonumber \\
\left\langle \widehat{a}_{i}^{\dagger}\widehat{a}_{j}^{\dagger}\right\rangle  & = & \left\langle \alpha_{i}^{+}\alpha_{j}^{+}+m_{ij}^{+}\right\rangle _{P}\,\,.\label{moments}\end{eqnarray}

}}

~

Comparing these equations with the schematic diagram in Fig \ref{figure1},
we see that, as expected from a convolution, the overall variance
of any quantity is the sum of the variances of the two convolved distributions;
that is, $\sigma=\sigma_{\Lambda}+\sigma_{P}$. The results also support
our interpretation given in section \ref{sub:Extended-phase-space}
that $\mathbf{n}$ and $\mathbf{m}$ are, respectively, the normal
and anomalous correlations that appear in many-body theory - except
for the additional feature that we can now allow for distributions
over these correlations. The expressions in the $P$-averages on the
right-hand side are not complex-conjugate for Hermitian-conjugate
operators, because the kernel $\widehat{\Lambda}(\overrightarrow{\lambda})$
is generically not Hermitian. Of course, after averaging over the
entire distribution, one must recover an Hermitian density matrix,
and hence the final expectation values of annihilation and creation
operators will be complex-conjugate. Using the characteristic function,
one can extend these to higher order moments via the standard Gaussian
factorizations in which odd moments of fluctuations vanish, and even
moments of fluctuations are expressed as the sum over all possible
distinct pairwise correlations.

\subsection{Quantum field expectation values}

The results obtained above can be applied directly to obtaining the
corresponding expectation values of normally ordered field operators:

\begin{eqnarray}
\left\langle \widehat{\Psi}_{i}(\mathbf{x})\right\rangle  & = & \left\langle \Psi_{i}(\mathbf{x})\right\rangle _{P}\nonumber \\
\left\langle \widehat{\Psi}_{i}^{\dagger}(\mathbf{x})\right\rangle  & = & \left\langle \Psi_{i}^{+}(\mathbf{x})\right\rangle _{P}\,\nonumber \\
\left\langle \widehat{\Psi}_{i}(\mathbf{x})\widehat{\Psi}_{j}(\mathbf{y})\right\rangle  & = & \left\langle \Psi_{i}(\mathbf{x})\Psi_{j}(\mathbf{y})+m_{ij}(\mathbf{x},\mathbf{y})\right\rangle _{P}\nonumber \\
\left\langle :\widehat{\Psi}_{i}(\mathbf{x})\widehat{\Psi}_{j}^{\dagger}(\mathbf{y}):\right\rangle  & = & \left\langle \Psi_{i}(\mathbf{x})\Psi_{j}^{+}(\mathbf{y})+n_{ij}(\mathbf{x},\mathbf{y})\right\rangle _{P}\,\nonumber \\
\left\langle \widehat{\Psi}_{i}^{\dagger}(\mathbf{x})\widehat{\Psi}_{j}^{\dagger}(\mathbf{y})\right\rangle  & = & \left\langle \Psi_{i}^{+}(\mathbf{x})\Psi_{j}^{+}(\mathbf{y})+m_{ij}^{+}(\mathbf{x},\mathbf{y})\right\rangle _{P}\,\,.\nonumber \\
\end{eqnarray}
These results show that in the field formulation of the Gaussian representation,
the phase-space quantities $n_{ij}(\mathbf{x},\mathbf{y})$ and $m_{ij}(\mathbf{x},\mathbf{y})$
correspond to single-time Greens functions, analogous to those found
in the propagator theory of quantum fields.

\subsection{Comparisons with other methods}

It is useful at this stage to compare these operator correspondences
with the most commonly used previously known representations, as shown
in Table \ref{cap:Representation}. For simplicity, this table only
gives a single-mode comparison:%
\begin{table}
\begin{tabular}{|c|c|c|c|c|c|c|}
\hline 
Representation&
$\Omega$&
$\alpha$&
$\alpha^{+}$&
$n$&
$m$&
$m^{+}$\tabularnewline
\hline
\hline 
Wigner ($W$)\cite{Wig-Wigner}&
$1$&
$\alpha$&
$\alpha^{\ast}$&
-$\frac{1}{2}$&
$0$&
$0$\tabularnewline
\hline 
Husimi ($Q$)\cite{Hus-Q}&
$1$&
$\alpha$&
$\alpha^{\ast}$&
$-1$&
$0$&
$0$\tabularnewline
\hline 
Glauber-Sudarshan ($P$)\cite{Gla-P}&
$1$&
$\alpha$&
$\alpha^{\ast}$&
$0$&
$0$&
$0$\tabularnewline
\hline 
$s$-ordered \cite{CG-Q,Agarwal}&
$1$&
$\alpha$&
$\alpha^{*}$&
$(s-1)/2$&
$0$&
$0$\tabularnewline
\hline 
squeezed \cite{CG-Q,SchCav84}&
$1$&
$\alpha$&
$\alpha^{*}$&
$n(|m|)$&
$m$&
$m^{\ast}$\tabularnewline
\hline 
Drummond-Gardiner ($+P$)\cite{CG-Q}&
$1$&
$\alpha$&
$\alpha^{+}$&
$0$&
$0$&
$0$\tabularnewline
\hline 
stochastic gauge\cite{gauge_paper,Paris1}&
$\Omega$&
$\alpha$&
$\alpha^{+}$&
$0$&
$0$&
$0$\tabularnewline
\hline
\end{tabular}

\caption{Classification of commonly used single-mode operator representations
in terms of parameters of the general Gaussian basis.}

\label{cap:Representation}
\end{table}

In greater detail, we notice that

\begin{itemize}
\item If $\sigma_{\mu\nu}=\delta_{\mu\nu}$, these results correspond to
the standard ones for the normally ordered positive-$P$ representation. 
\item If we consider the Hermitian case of $\alpha^{\ast}=\alpha^{+}$ as
well, but with $\sigma_{\mu\nu}=(n-1)\delta_{\mu\nu}$, where $n=(s-1)/2$,
we obtain the `$s$-ordered' representation correspondences of Cahill
and Glauber. 
\item These include, as special cases, the normally ordered Glauber-Sudarshan
$P$ representation ($n=0$), and the symmetrically ordered representation
of Wigner ($n=-1/2$)
\item The antinormally ordered Husimi $Q$ function is recovered as the
singular limit $n\rightarrow-1$ . 
\item In the squeezed-state basis, the parameters $n$, $m$ are not independent,
as indicated in the table. The particle number $n$ is a function
$n(|m|)$of the squeezing $m$. The exact relationship is given later.
\item The Gaussian family of representations is much larger than the traditional
phase-space variety, because we can allow other values of the $\sigma_{\mu\nu}$
variance -- for example, squeezed or thermal state bases. For thermal
states, the variance corresponds to a Hermitian, positive-definite
density matrix if $n_{ij}$ is Hermitian and positive definite, in
which case $n_{ij}$ behaves analogously to the Green's function in
a bosonic field theory. In this case, a unitary transformation of
the operators can always be used to diagonalize $n_{ij}$, so that
$n_{ij}=n_{i}\delta_{ij}$. 
\item For a general Gaussian basis, Gaussian operators that do not themselves
satisfy density matrix requirements are permitted as part of the basis
-- provided the distribution has a finite width to compensate for
this. This is precisely what happens, for example, with the well-known
$Q$ function, which always has a positive variance to compensate
for the lack of fluctuations in the corresponding basis, which is
Hermitian but not positive-definite.
\end{itemize}
Distributions over the variance are also possible. It is the introduction
of distributions over the variance that represents the most drastic
change from the older distribution methods. It means that there many
new operator correspondences to use. Thus, the covariance itself can
be introduced as a dynamical variable in phase space, which can change
and fluctuate with time. In this respect, the present methods have
a similarity with the Kohn variational technique, which uses a density
in coordinate space, and has been suggested in the context of BEC\cite{Kohn}.
Related variational methods using squeezed states have also been utilized
for BEC problems\cite{Navez}. By comparison, the present methods
do not require either the local density approximation or variational
approximations.

\section{Gaussian differential identities}

\label{sec:Gaussian-differential-identities}An important application
of phase-space representations is to simulate canonical ensembles
and quantum dynamics in a phase space. An essential step in this process
is to map the master equation of a quantum density operator onto a
Liouville equation for the probability distribution $P$. The real
or imaginary time evolution of a quantum system depends on the action
of Hamiltonian operators on the density matrix. Thus it is useful
to have identities that describe the action of any quadratic bosonic
form as derivatives on elements of the Gaussian basis. These derivatives
can, by integration by parts, be applied to the distribution $P$,
provided boundary terms vanish. The resultant Liouville equation for
$P$ is equivalent to the original master equation, given certain
restrictions on the radial growth of the distribution. When the Liouville
equation has derivatives of only second order or less (and thus is
in the form of a Fokker-Planck equation), it is possible to obtain
an equivalent stochastic differential equation which can be efficiently
simulated.

In general, there are many ways to obtain these identities, but we
are interested in identities which result in first-order derivatives,
where possible. Just as for expectation values, this can be achieved
most readily if the operator $\widehat{O}$ is written in factorized
form, as in Eq (\ref{eq:anti-norm}).

In this notation, normal ordering means:

\begin{equation}
:\widehat{O}\widehat{\Lambda}:=\sum_{i}\widehat{v}_{i}(\widehat{\bm a}^{\dagger})\widehat{\Lambda}\widehat{u}_{i}(\widehat{\bm a})\,\,.\label{eq:normal}\end{equation}
We also need a notation for partial antinormal ordering:\begin{equation}
\left\{ \widehat{O}:\widehat{\Lambda}:\right\} =\sum_{i}\widehat{u}_{i}(\widehat{\bm a}):\widehat{\Lambda}:\widehat{v}_{i}(\widehat{\bm a}^{\dagger})\,\,,\label{eq:partial-anti-norm}\end{equation}
which indicates an operator product which antinormally orders all
terms except the normal term $:\widehat{\Lambda}:$. The Gaussian
kernel $\widehat{\Lambda}$ is always normally ordered, and hence
we can omit the explicit normal-ordering notation, without ambiguity,
for the kernel itself.

In this section, for brevity, we use ${\partial/\partial\overrightarrow{\lambda}}=(\partial/\Omega,\partial/\bm\alpha,\partial/\bm\alpha^{+},\partial/\mathbf{n},\partial/\mathbf{m},\partial/\mathbf{m}^{+})$
to symbolize either $\partial/\partial x_{i}$ or $-i\partial/\partial y_{i}$
for each of the $i=0,..\, p$ complex variables $\overrightarrow{\lambda}$
. This is possible since $\widehat{\Lambda}(\overrightarrow{\lambda})$
is an analytic function of $\overrightarrow{\lambda}$ , and an explicit
choice of derivative can be made later. We first note a trivial identity,
which is nevertheless useful in obtaining stochastic gauge equivalences
between the different possible forms of time-evolution equations:\begin{equation}
\Omega\frac{\partial}{\partial\Omega}\widehat{\Lambda}=\widehat{\Lambda}\,\,.\label{eq:linear0}\end{equation}

\subsection{Normally ordered identities}

The normally ordered operator product identities can be calculated
simply by taking a derivative of the Gaussian operator with respect
to the amplitude and variance parameters.

\subsubsection{Linear products: }

The result for linear operator products follows directly from differentiation
with respect to the coherent amplitude, noting that each amplitude
appears twice in the exponent:

\begin{eqnarray}
\frac{\partial}{\partial\alpha_{\mu}^{+}}\widehat{\Lambda} & = & \frac{\partial}{\partial\alpha_{\mu}^{+}}\frac{\Omega}{\sqrt{\left|\underline{\underline{\sigma}}\right|}}:\exp\left[-\delta\underline{\widehat{a}}^{\dagger}\underline{\underline{\sigma}}^{-1}\delta\underline{\widehat{a}}/2\right]:\nonumber \\
 & = & [\sigma^{-1}]_{\mu\nu}:\delta\widehat{a}_{\nu}\widehat{\Lambda}:\,\,.\label{eq:linear1}\end{eqnarray}
It follows that:\begin{equation}
:\widehat{a}_{\mu}\widehat{\Lambda}:=\left[\alpha_{\mu}+\sigma_{\mu\nu}\frac{\partial}{\partial\alpha_{\nu}^{+}}\right]\widehat{\Lambda}\,\,.\label{eq:linear2}\end{equation}

\subsubsection{Quadratic products: }

Differentiating a determinant results in a transposed inverse, a result
that follows from the standard cofactor expansion of determinants:\begin{equation}
\frac{\partial\left|\underline{\underline{\sigma}}\right|}{\partial\sigma_{\nu\mu}}=\sigma_{\mu\nu}^{-1}\left|\underline{\underline{\sigma}}\right|\,\,.\label{eq:detderiv1}\end{equation}
 Similarly, for the normalization factor that occurs in Gaussian operators:\begin{equation}
\frac{\partial\left|\underline{\underline{\sigma}}\right|^{-\frac{1}{2}}}{\partial\sigma_{\nu\mu}^{-1}}=\frac{1}{2}\sigma_{\mu\nu}\left|\underline{\underline{\sigma}}\right|^{-\frac{1}{2}}\,\,.\label{eq:detderiv2}\end{equation}
Hence, on differentiating with respect to the inverse covariance,
we can obtain the following identity for any normalized Gaussian operator:\\
\begin{eqnarray}
\frac{\partial}{\partial\sigma_{\nu\mu}^{-1}}\widehat{\Lambda} & = & \frac{\partial}{\partial\sigma_{\nu\mu}^{-1}}\frac{\Omega}{\sqrt{\left|\underline{\underline{\sigma}}\right|}}:\exp\left[-\delta\underline{\widehat{a}}^{\dagger}\cdot\underline{\underline{\sigma}}^{-1}\cdot\delta\underline{\widehat{a}}/2\right]:\nonumber \\
 & = & \frac{1}{2}:\left[\sigma_{\mu\nu}-\delta\widehat{a}_{\mu}\delta\widehat{a}_{\nu}^{\dagger}\right]\widehat{\Lambda}:\,\,.\label{eq:Quad1}\end{eqnarray}
Using the chain rule to transform the derivative, it follows that
a normally ordered quadratic product has the following identity:\begin{eqnarray}
:\delta\widehat{a}_{\mu}\delta\widehat{a}_{\nu}^{\dagger}\widehat{\Lambda}: & = & \left[\sigma_{\mu\nu}-2\frac{\partial}{\partial\sigma_{\nu\mu}^{-1}}\right]\widehat{\Lambda}\,\,\,\,\nonumber \\
 & = & \left[\sigma_{\mu\nu}+2\sigma_{\mu\alpha}\sigma_{\beta\nu}\frac{\partial}{\partial\sigma_{\beta\alpha}}\right]\widehat{\Lambda}\,\,.\label{eq:Quad2}\end{eqnarray}

\subsection{Antinormally ordered identities}

The antinormally ordered operator product identities are all obtained
from the above results, on making use of the algebraic re-ordering
results in the Appendix:

\subsubsection{Antinormal linear products}

Antinormally ordered linear products can be transformed directly to
normally ordered products. Hence, from the Appendix and Eq (\ref{eq:linear2}),
we obtain:\begin{eqnarray}
\left\{ \widehat{a}_{\mu}:\widehat{\Lambda}:\right\}  & = & :\left[\widehat{a}_{\mu}-\sigma_{\mu\nu}^{-1}\delta\widehat{a}_{\nu}\right]\widehat{\Lambda}:\nonumber \\
 & = & \left[\alpha_{\nu}+(\sigma_{\mu\nu}-\delta_{\mu\nu})\frac{\partial}{\partial\alpha_{\nu}^{+}}\right]\widehat{\Lambda}\nonumber \\
 & = & \left[\alpha_{\nu}+\sigma_{\mu\nu}^{N}\frac{\partial}{\partial\alpha_{\nu}^{+}}\right]\widehat{\Lambda}\,\,,\end{eqnarray}
where we recall from Eq (\ref{eq:normalcov}) that the normally ordered
covariance is defined by: $\underline{\underline{\sigma}}^{N}=\underline{\underline{\sigma}}-\underline{\underline{I}}$.

\subsubsection{Quadratic products with one antinormal operator}

This calculation follows a similar pattern to the previous one:

\begin{eqnarray}
\left\{ \delta\widehat{a}_{\mu}:\delta\widehat{a}_{\nu}^{\dagger}\widehat{\Lambda}:\right\}  & = & :\left[\delta\widehat{a}_{\mu}+\frac{\partial}{\partial\widehat{a}_{\mu}^{\dagger}}\right]:\delta\widehat{a}_{\nu}^{\dagger}\widehat{\Lambda}:\nonumber \\
 & = & :\left[\delta_{\mu\nu}+(\delta_{\mu\rho}-\sigma_{\mu\rho}^{-1})\delta\widehat{a}_{\nu}^{\dagger}\delta\widehat{a}_{\rho}\right]\widehat{\Lambda}:\nonumber \\
 & = & \left[\sigma_{\mu\nu}+2\sigma_{\mu\alpha}^{N}\sigma_{\beta\nu}\frac{\partial}{\partial\sigma_{\beta\alpha}}\right]\widehat{\Lambda}\,\,.\label{eq:Quad1a}\end{eqnarray}

\subsubsection{Quadratic products with two antinormal operators}

We first expand this as the iterated result of two re-orderings, then
apply the result for a linear antinormal product to the innermost
bracket:\begin{eqnarray}
\left\{ \delta\widehat{a}_{\mu}\delta\widehat{a}_{\nu}^{\dagger}:\widehat{\Lambda}:\right\}  & = & \left\{ \delta\widehat{a}_{\mu}\left\{ \delta\widehat{a}_{\nu}^{\dagger}:\widehat{\Lambda}:\right\} \right\} \nonumber \\
 & = & \left\{ \delta\widehat{a}_{\mu}\left[\delta_{\nu\rho}-\sigma_{\rho\nu}^{-1}\right]:\delta\widehat{a}_{\rho}^{\dagger}\widehat{\Lambda}:\right\} \,\,.\label{eq:Quad2a}\end{eqnarray}
 Next, the result above for one antinormal operator is used:\begin{eqnarray}
\left\{ \delta\widehat{a}_{\mu}\delta\widehat{a}_{\nu}^{\dagger}:\widehat{\Lambda}:\right\}  & = & \left[\delta_{\nu\rho}-\sigma_{\rho\nu}^{-1}\right]\nonumber \\
 & \times & \left[\sigma_{\mu\rho}+2(\sigma_{\mu\alpha}-\delta_{\mu\alpha})\sigma_{\beta\rho}\frac{\partial}{\partial\sigma_{\beta\alpha}}\right]\widehat{\Lambda}\nonumber \\
 & = & \left[\sigma_{\mu\nu}^{N}+2\sigma_{\mu\alpha}^{N}\sigma_{\beta\nu}^{N}\frac{\partial}{\partial\sigma_{\beta\alpha}}\right]\widehat{\Lambda}\,\,.\label{eq:Quad2b}\end{eqnarray}

\subsection{Identities in matrix form}

The different possible quadratic orderings can be written in matrix
form as

\begin{eqnarray}
:\widehat{\underline{{a}}}\,\widehat{\underline{{a}}}^{\dagger}\widehat{\Lambda}: & = & \left[\begin{array}{cc}
:\widehat{\bm a}\widehat{\bm a}^{\dagger}\widehat{\Lambda}: & \widehat{\Lambda}\widehat{\bm a}\widehat{\bm a}^{T}\\
\widehat{\bm a}^{\dagger T}\widehat{\bm a}^{\dagger}\widehat{\Lambda} & \widehat{\bm a}^{\dagger T}\widehat{\Lambda}\widehat{\bm a}^{T}\end{array}\right]\,\,,\nonumber \\
\left\{ \widehat{\underline{{a}}}\,\widehat{\underline{{a}}}^{\dagger}\widehat{\Lambda}\right\}  & = & \left[\begin{array}{cc}
\widehat{\bm a}\widehat{\Lambda}\widehat{\bm a}^{\dagger} & \widehat{\bm a}\widehat{\bm a}^{T}\widehat{\Lambda}\\
\widehat{\Lambda}\widehat{\bm a}^{\dagger T}\widehat{\bm a}^{\dagger} & \left\{ \widehat{\bm a}^{\dagger T}\widehat{\Lambda}\widehat{\bm a}^{T}\right\} \end{array}\right]\,\,,\nonumber \\
\left\{ \widehat{\underline{{a}}}\,:\widehat{\underline{{a}}}^{\dagger}\widehat{\Lambda}:\right\}  & = & \left[\begin{array}{cc}
\widehat{\bm a}\widehat{\bm a}^{\dagger}\widehat{\Lambda} & \widehat{\bm a}\widehat{\Lambda}\widehat{\bm a}^{T}\\
\widehat{\bm a}^{\dagger T}\widehat{\Lambda}\widehat{\bm a}^{\dagger} & \left\{ \widehat{\bm a}^{\dagger T}:\widehat{\bm a}^{T}\widehat{\Lambda}:\right\} \end{array}\right]\,\,.\label{eq:Matrixform}\end{eqnarray}

With this notation, all of the operator identities can be written
in a compact matrix form. The resulting set of differential identities
can be used to map any possible linear or quadratic operator acting
on the kernel $\widehat{\Lambda}$ into a first-order differential
operator acting on the kernel. 

For this reason, the following identities are the central result of
this paper:\\

\fbox{\parbox{0.8\columnwidth}{

\begin{eqnarray}
\widehat{\Lambda} & = & \Omega\frac{\partial}{\partial\Omega}\widehat{\Lambda}\nonumber \\
:\widehat{\underline{a}}\widehat{\Lambda}: & = & \underline{\alpha}\widehat{\Lambda}+\underline{\underline{\sigma}}\frac{\partial\widehat{\Lambda}}{\partial\underline{{\alpha}}^{+}},\nonumber \\
\left\{ \widehat{\underline{a}}\widehat{\Lambda}\right\}  & = & \underline{\alpha}\widehat{\Lambda}+\underline{\underline{\sigma}}^{N}\frac{\partial\widehat{\Lambda}}{\partial\underline{\alpha}^{+}},\nonumber \\
:\delta\widehat{\underline{a}}\delta\widehat{\underline{a}}^{\dagger}\widehat{\Lambda}: & = & \underline{\underline{\sigma}}\widehat{\Lambda}+2\underline{\underline{\sigma}}\frac{\partial\widehat{\Lambda}}{\partial\underline{\underline{\sigma}}}\underline{\underline{\sigma}}\,\,,\nonumber \\
\left\{ \delta\widehat{\underline{a}}:\delta\widehat{\underline{a}}^{\dagger}\widehat{\Lambda}:\right\}  & = & \underline{\underline{\sigma}}\widehat{\Lambda}+2\underline{\underline{\sigma}}^{N}\frac{\partial\widehat{\Lambda}}{\partial\underline{\underline{\sigma}}}\underline{\underline{\sigma}}\,\,,\nonumber \\
\left\{ \delta\widehat{\underline{a}}\delta\widehat{\underline{a}}^{\dagger}\widehat{\Lambda}\right\}  & = & \underline{\underline{\sigma}}^{N}\widehat{\Lambda}+2\underline{\underline{\sigma}}^{N}\frac{\partial\widehat{\Lambda}}{\partial\underline{\underline{\sigma}}}\underline{\underline{\sigma}}^{N}\,\,.\nonumber \\
\label{eq:Matrixidentities}\end{eqnarray}
 }} 

~

Here the derivatives are defined as \begin{eqnarray}
\frac{\partial}{\partial\underline{\alpha}} & = & \left(\frac{\partial}{\partial\bm\alpha},\left(\frac{\partial}{\partial\bm\alpha^{+}}\right)^{T}\right)\nonumber \\
\frac{\partial}{\partial\underline{\alpha}^{+}} & = & \left(\begin{array}{c}
\partial/\partial\bm\alpha^{+}\\
\left(\partial/\partial\bm\alpha\right)^{T}\end{array}\right)\nonumber \\
\left(\frac{\partial}{\partial\underline{\underline{\sigma}}}\right)_{\mu,\nu} & = & \frac{\partial}{\partial\sigma_{\nu\mu}}\,\,.\label{eq:definederiv}\end{eqnarray}
\\
It should be noted that the matrix and vector derivatives involve
taking the transpose. We note here that for notational convenience,
the derivatives with respect to the $\sigma_{\mu,\nu}$ are formal
derivatives, calculated as if each of the $\sigma_{\nu,\mu}$ were
independent of the others. With a symmetry constraint, the actual
derivatives of $\widehat{\Lambda}$ with respect to any elements of
$\bm n$ or any off-diagonal elements of $\bm m$ will differ from
the formal derivatives by a factor of two. Fortunately, because of
the summation over all derivatives in the final Fokker-Planck equation,
the final results are the same, regardless of whether or not the symmetry
of $\sigma_{\mu,\nu}$ is explicitly taken into account at this stage. 

The quadratic terms can also be written in a form without the coherent
offset terms in the operator products. This is often useful, since
while the original Hamiltonian or master-equation may not have an
explicit coherent term, terms like this can arise dynamically. The
following result is obtained:\begin{eqnarray}
:\widehat{\underline{a}}\,\widehat{\underline{a}}^{\dagger}\widehat{\Lambda}: & = & \underline{\alpha}\frac{\partial\widehat{\Lambda}}{\partial\underline{\alpha}}\underline{\underline{\sigma}}+\underline{\underline{\sigma}}\frac{\partial\widehat{\Lambda}}{\partial\underline{\alpha}^{+}}\underline{\alpha}^{+}\nonumber \\
 & + & \left(\widehat{\underline{\alpha}}\,\widehat{\underline{\alpha}}^{+}+\underline{\underline{\sigma}}\right)\widehat{\Lambda}+2\underline{\underline{\sigma}}\frac{\partial\widehat{\Lambda}}{\partial\underline{\underline{\sigma}}}\underline{\underline{\sigma}},\nonumber \\
\left\{ \widehat{\underline{a}}\,:\widehat{\underline{a}}^{\dagger}\widehat{\Lambda}:\right\}  & = & \underline{\alpha}\frac{\partial\widehat{\Lambda}}{\partial\underline{\alpha}}\underline{\underline{\sigma}}+\underline{\underline{\sigma}}^{N}\frac{\partial\widehat{\Lambda}}{\partial\underline{\alpha}^{+}}\underline{\alpha}^{+}\nonumber \\
 & + & \left(\widehat{\underline{\alpha}}\,\widehat{\underline{\alpha}}^{+}+\underline{\underline{\sigma}}\right)\widehat{\Lambda}+2\underline{\underline{\sigma}}^{N}\frac{\partial\widehat{\Lambda}}{\partial\underline{\underline{\sigma}}}\underline{\underline{\sigma}},\nonumber \\
\left\{ \widehat{\underline{a}}\,\widehat{\underline{a}}^{\dagger}\widehat{\Lambda}\right\}  & = & \underline{\alpha}\frac{\partial\widehat{\Lambda}}{\partial\underline{\alpha}}\underline{\underline{\sigma}}^{N}+\underline{\underline{\sigma}}^{N}\frac{\partial\widehat{\Lambda}}{\partial\underline{\alpha}^{+}}\underline{\alpha}^{+}\nonumber \\
 & + & \left(\widehat{\underline{\alpha}}\,\widehat{\underline{\alpha}}^{+}+\underline{\underline{\sigma}}^{N}\right)\widehat{\Lambda}+2\underline{\underline{\sigma}}^{N}\frac{\partial\widehat{\Lambda}}{\partial\underline{\underline{\sigma}}}\underline{\underline{\sigma}}^{N}\,\,.\label{eq:nocoherent}\end{eqnarray}

One consequence of these identities is that the time evolution resulting
from a quadratic Hamiltonian can always be expressed as a simple first-order
differential equation, which therefore corresponds to a deterministic
trajectory. This relationship will be explored in later sections:
it is quite different to the result of a path integral, which gives
a sum over many fluctuating paths for a quadratic Hamiltonian. Similarly,
the time evolution for cubic and quartic Hamiltonians can always be
expressed as a second-order differential equation, which corresponds
to a stochastic trajectory.

\subsection{Identities for quantum field operators}

The operator mappings can also be succinctly written in the field-theoretic
notation as\begin{eqnarray}
:\widehat{\underline{\Psi}}(\mathbf{x})\widehat{\Lambda}: & = & \underline{\Psi}(\mathbf{x})\widehat{\Lambda}+\int\!\! d^{3}\! x'\underline{\underline{\sigma}}(\mathbf{x},\mathbf{x}')\frac{\partial\widehat{\Lambda}}{\partial\underline{\Psi}^{+}(\mathbf{x}')}\nonumber \\
\left\{ \widehat{\underline{\Psi}}(\mathbf{x})\widehat{\Lambda}\right\}  & = & \underline{\Psi}(\mathbf{x})\widehat{\Lambda}+\int\!\! d^{3}\! x'\underline{\underline{\sigma}}^{N}(\mathbf{x},\mathbf{x}')\frac{\partial\widehat{\Lambda}}{\partial\underline{\Psi}^{+}(\mathbf{x}')}\nonumber \\
:\delta\widehat{\underline{\Psi}}(\mathbf{x})\delta\widehat{\underline{\Psi}}(\mathbf{x'})^{\dagger}\widehat{\Lambda}: & = & \underline{\underline{\sigma}}(\mathbf{x},\mathbf{x}')\widehat{\Lambda}+2\int\!\!\int\!\! d^{3}\! x''d^{3}\! x'''\times\nonumber \\
 &  & \underline{\underline{\sigma}}(\mathbf{x},\mathbf{x}'')\frac{\partial\widehat{\Lambda}}{\partial\underline{\underline{\sigma}}(\mathbf{x}'',\mathbf{x}''')}\underline{\underline{\sigma}}(\mathbf{x}''',\mathbf{x}')\nonumber \\
\left\{ \delta\widehat{\underline{\Psi}}(\mathbf{x}):\delta\widehat{\underline{\Psi}}(\mathbf{x'})^{\dagger}\widehat{\Lambda}:\right\}  & = & \underline{\underline{\sigma}}(\mathbf{x},\mathbf{x}')\widehat{\Lambda}+2\int\!\!\int\!\! d^{3}\! x''d^{3}\! x'''\times\nonumber \\
 &  & \underline{\underline{\sigma}}^{N}(\mathbf{x},\mathbf{x}'')\frac{\partial\widehat{\Lambda}}{\partial\underline{\underline{\sigma}}(\mathbf{x}'',\mathbf{x}''')}\underline{\underline{\sigma}}(\mathbf{x}''',\mathbf{x}')\nonumber \\
\left\{ \delta\widehat{\underline{\Psi}}(\mathbf{x})\delta\widehat{\underline{\Psi}}(\mathbf{x'})^{\dagger}\widehat{\Lambda}\right\}  & = & \underline{\underline{\sigma}}^{N}(\mathbf{x},\mathbf{x}')\widehat{\Lambda}+2\int\!\!\int\!\! d^{3}\! x''d^{3}\! x'''\times\nonumber \\
 &  & \underline{\underline{\sigma}}^{N}(\mathbf{x},\mathbf{x}'')\frac{\partial\widehat{\Lambda}}{\partial\underline{\underline{\sigma}}(\mathbf{x}'',\mathbf{x}''')}\underline{\underline{\sigma}^{N}}(\mathbf{x}''',\mathbf{x}'),\nonumber \\
\label{eq:Fieldidentities}\end{eqnarray}
where the vector quantum fields and covariances are as defined in
section \ref{sub:Gaussian-field-operators}. The normal field correlation
matrix is $\underline{\underline{\sigma}}^{N}(\mathbf{x},\mathbf{x}'')=\underline{\underline{\sigma}}(\mathbf{x},\mathbf{x}'')-\underline{\underline{I}}\delta(\mathbf{x},\mathbf{x}')$
and the functional derivatives have been defined as\begin{eqnarray*}
\frac{\partial}{\partial\Psi_{js}(\mathbf{x})} & = & \frac{1}{\sqrt{V}}\sum_{\mathbf{k}}e^{-is\mathbf{k}.\mathbf{x}}\frac{\partial}{\partial\alpha_{\mathbf{k}js}},\\
\frac{\partial}{\partial\Psi_{js}^{+}(\mathbf{x})} & = & \frac{1}{\sqrt{V}}\sum_{\mathbf{k}}e^{is\mathbf{k}.\mathbf{x}}\frac{\partial}{\partial\alpha_{\mathbf{k}js}^{+}},\\
\frac{\partial}{\partial\sigma_{js,j's'}(\mathbf{x},\mathbf{x}')} & = & \frac{1}{V}\sum_{\mathbf{k}}\sum_{\mathbf{k}'}e^{-i(s'\mathbf{k}\cdot\mathbf{x}-s\mathbf{k}'\cdot\mathbf{x}')}\frac{\partial}{\partial\sigma_{\mathbf{k}'js,\mathbf{k}j's'}}\,.\end{eqnarray*}
 Again we have the convention for matrix derivatives that \begin{eqnarray*}
\left(\frac{\partial}{\partial\underline{\underline{\sigma}}(\mathbf{x},\mathbf{x'})}\right)_{js,j's'} & = & \frac{\partial}{\partial\sigma_{j's',js}(\mathbf{x},\mathbf{x}')}\,.\end{eqnarray*}

\section{Examples of Gaussian operators}

\label{sec:Examples}This section focuses on specific examples of
Gaussian operators, and relates them to physically useful pure states
or density matrices. We begin by defining the class of Gaussian operators
that correspond to physical density matrices, before looking at examples
of specific types of states that can be represented, such as coherent,
squeezed and thermal. In each of these specific cases, the conventional
parametrization can be analytically continued to describe a non-Hermitian
basis for a positive representation. We show how these bases include
and extend those of previously defined representations, and calculate
the normalization rules and identities that apply in the simpler cases.

\subsection{Gaussian density matrices}

A Gaussian operator can itself correspond to a physical density matrix,
in which case the corresponding distribution is a delta function.
This is the simplest possible representation of a physical state.
Gaussian states or physical density matrices are required to satisfy
the usual constraints necessary for any density matrix: they must
be Hermitian and positive definite. From the moment results of Eq
(\ref{moments}), the requirement of Hermiticity generates the following
immediate restrictions on the displacement and covariance parameters:

\begin{eqnarray}
\bm\alpha^{\dagger} & = & \bm\alpha^{+}\nonumber \\
\mathbf{n}^{\dagger} & = & \mathbf{n}\nonumber \\
\mathbf{m}^{\dagger} & = & \mathbf{m}^{+}\,\,.\label{eq:Hermitian-parameters}\end{eqnarray}

In addition, there are requirements due to positive-definiteness.
To understand these, we first note that when $\mathbf{n}$ is Hermitian,
as it must be for a density matrix, it is diagonalizable via a unitary
transformation on the mode operators. Therefore, with no loss of generality,
we can consider the case of diagonal $\mathbf{n}$, i.e., $n_{kj}=n_{k}\delta_{kj}$.
The positive-definiteness of the number operator then means that the
number eigenvalues are real and non-negative:\begin{equation}
n_{k}\ge0\,\,.\label{eq:n-pos}\end{equation}
In the diagonal thermal density matrix case, but with squeezed correlations
as well, satisfying the density matrix requirements means that there
are additional restrictions\cite{Squeezing_limits}. Consideration
of the positivity of products like $\widehat{X}_{kj}\widehat{X}_{kj}^{\dagger}$
where: $\widehat{X}_{k}=\mu\widehat{a}_{k}+\nu\widehat{a}_{j}^{\dagger}$
means that one must also satisfy the inequalities $n_{k}(1+n_{j})\ge\left|m_{kj}\right|^{2}$.
This implies a necessary lower bound on the photon number in each
mode:\begin{equation}
n_{k}\ge n(|m_{kk}|)=\sqrt{|m_{kk}|^{2}+1/4}-1/2\,\,.\label{eq:n-bound}\end{equation}

Examples of Gaussians of this type are readily obtained by first generating
a thermal density matrix, then applying unitary squeezing and/or coherent
displacement operations, which preserve the positive definite nature
of the original thermal state. This produces a pure state if and only
if the starting point is a zero-temperature thermal state or vacuum
state. Hence, the general physical density matrix can be written in
factorized form as:\begin{equation}
\widehat{\Lambda}_{\rho}=\widehat{{D}}(\bm\alpha)\widehat{S}(\bm\xi)\widehat{\Lambda}_{\textrm{th}}(\overline{\bm n})\widehat{S}(-\bm\xi)\widehat{{D}}(-\bm\alpha)\,\,.\label{eq:physicalcase}\end{equation}
Here \begin{equation}
\widehat{\Lambda}_{\textrm{th}}(\overline{\bm n})=\frac{1}{\left|\bm1+\overline{\bm n}\right|}:\exp\left[-\widehat{\bm a}^{\dagger}\left[\bm1+\overline{\bm n}\right]^{-1}\widehat{\bm a}\right]:\label{eq:thermalcase}\end{equation}
 is a thermal density matrix completely characterized by its number
expectation: $\overline{\bm n}\equiv{\rm {Tr}}\left[:\widehat{\bm a}\widehat{\bm a}^{\dagger}:\widehat{\Lambda}_{th}(\overline{\bm n})\right]$,
where $\overline{\bm n}$ must be Hermitian for the operator to correspond
to a physical density matrix. We show the equivalence of this expression
to the more standard canonical Bose-Einstein form in the next section.

The unitary displacement and squeezing operators are as usually defined
in the literature:\begin{equation}
\widehat{{D}}(\bm\alpha)=e^{\bm\alpha\widehat{\bm a}^{\dagger}-\widehat{\bm a}\bm\alpha^{*}}\end{equation}
and:\begin{equation}
\widehat{S}(\bm\xi)=e^{-\widehat{\bm a}^{\dagger}\bm\xi\widehat{\bm a}^{\dagger}/2+\widehat{\bm a}\bm\xi^{*}\widehat{\bm a}/2}\,\,,\end{equation}
 where the vector $\bm\alpha$ is, as before, the coherent displacements
for each mode. The symmetric matrix $\bm\xi$ gives the angle and
degree of squeezing for each mode, as well as the squeezing correlations
between each pair of modes.

In table (\ref{cap:Gaussian-parameters}), we give a comparison of
the Gaussian parameters found in the usual classifications of physical
density matrices of bosons, for a single-mode case.

\begin{table}
\begin{tabular}{|c|c|c|c|c|c|c|}
\hline 
Physical state&
$\Omega$&
$\alpha$&
$\alpha^{+}$&
$n$&
$m$&
$m^{+}$\tabularnewline
\hline
\hline 
Vacuum state&
$1$&
$0$&
$0$&
$0$&
$0$&
$0$\tabularnewline
\hline 
Coherent state&
$1$&
$\alpha$&
$\alpha^{\ast}$&
$0$&
$0$&
$0$\tabularnewline
\hline 
Thermal &
$1$&
$0$&
$0$&
$n\geq0$&
$0$&
$0$\tabularnewline
\hline 
Squeezed vacuum&
$1$&
$0$&
$0$&
$n(|m|)$&
$m$&
$m^{\ast}$\tabularnewline
\hline 
Squeezed coherent&
$1$&
$\alpha$&
$\alpha^{\ast}$&
$n(|m|)$&
$m$&
$m^{\ast}$\tabularnewline
\hline 
Squeezed thermal&
$1$&
$\alpha$&
$\alpha^{\ast}$&
$n\geq n(|m|)$&
$m$&
$m^{\ast}$\tabularnewline
\hline
\end{tabular}

\caption{Parameters of single-mode Gaussian density matrices of bosons.\label{cap:Gaussian-parameters}}
\end{table}

\subsection{Thermal operators}

\subsubsection{Physical states}

It is conventional to write the bosonic thermal density operator for
a non-interacting Bose gas in grand canonical form as\cite{Louisell}:\begin{equation}
\widehat{\rho}_{\textrm{th}}(\bm\phi)=\prod_{k}\left[1-e^{-\phi_{k}}\right]\exp\left[-\phi_{k}\widehat{a}_{k}^{\dagger}\widehat{a}_{k}\right]\,\,,\label{Th_{u}nordered}\end{equation}
where $\phi_{k}=\epsilon_{k}/kT$ . Here the modes are chosen, with
no loss of generality, to diagonalize the free Hamiltonian with mode
energies $\epsilon_{k}$, and for the case of massive bosons we have
included the chemical potential in the definition of the energy origin.
To show how this form is related to the normally ordered thermal Gaussian
$\widehat{\Lambda}_{\textrm{th}}(\overline{\bm n})$ of Eq (\ref{eq:thermalcase})
, we simply note that since $\overline{\bm n}$ is hermitian, it can
be diagonalized by a unitary transformation. The resulting diagonal
form in either expression is therefore diagonal in a number state
basis and is uniquely defined by its number state expectation value.

Clearly, one has for the usual canonical density matrix that:

\begin{equation}
\left\langle \mathbf{n}\right|\widehat{\rho}_{\textrm{th}}\left|\mathbf{n}\right\rangle =\prod_{k}\left[1-e^{-\phi_{k}}\right]\exp\left[-\phi_{k}n_{k}\right]\,\,,\end{equation}
while it is straight-forward to show that the corresponding normally
ordered expression is a binomial: \begin{equation}
\left\langle \mathbf{n}\right|\widehat{\Lambda}_{\textrm{th}}(\overline{\bm n})\left|\mathbf{n}\right\rangle =\prod_{k}\left[1+\overline{n}_{k}\right]^{-1}\left[1-\frac{1}{1+\overline{n}_{k}}\right]^{n_{k}}\,\,.\end{equation}
As one would expect, these expressions are identical provided one
chooses the standard Bose-Einstein result for the thermal occupation
as:\begin{equation}
\overline{n}_{k}=\frac{1}{e^{\phi_{k}}-1}\,\,.\label{eq:BoseEinstein}\end{equation}

These results also show that when $\overline{\bm n}=0$ one has a
vacuum state, corresponding to a bosonic ground state at zero temperature.
In summary, the normally ordered thermal Gaussian state \emph{}is
completely equivalent to the usual canonical form.

\subsubsection{Generalized thermal operators}

A simple non-Hermitian extension of the thermal states can be defined
as an analytic continuation of the usual Bose-Einstein density matrix
for bosons in thermal equilibrium. We define a normally ordered thermal
Gaussian \emph{operator} as having zero mean displacement and zero
second or fourth quadrant variance:\begin{equation}
\widehat{\Lambda}(\Omega,\mathbf{0},\mathbf{0},\mathbf{n},\mathbf{0},\mathbf{0})=\frac{\Omega}{\left|\mathbf{I}+\mathbf{n}\right|}:\exp\left[(\mathbf{I}+\mathbf{n})_{ij}^{-1}\widehat{a}_{i}^{\dagger}\widehat{a}_{j}\right]:\,\,.\label{Th_{o}rdered}\end{equation}
Such operators are an analytic continuation of previously defined
thermal bases, and are related to the thermo-field methods\cite{ChaSriAga99}. 

As well as the usual Bose-Einstein thermal distribution, the extended
thermal basis can represent a variety of other physical states. As
an example, consider the general matrix elements of an analytically
continued single-mode thermal Gaussian operator in a number-state
basis, with $1+1/\overline{n}=\exp\left[\phi\right]=\exp\left[(r+i\psi)\right]$.
These are:

\begin{eqnarray}
\left\langle n\right|\widehat{\Lambda}_{\textrm{th}}(\bar{n}\,)\left|n'\right\rangle  & = & \left\langle n\right|\left[1-e^{-\phi}\right]\exp\left[-\phi\widehat{n}\right]\left|n'\right\rangle \nonumber \\
 & = & \delta_{\mathbf{nn'}}\left[1-e^{-\phi}\right]\exp\left[-n(r+i\psi)\right]\,\,.\nonumber \\
\label{eq:innerprod}\end{eqnarray}
Now consider the following single-mode density matrix:\begin{equation}
\widehat{\rho}=\frac{1}{2\pi}\int_{0}^{2\pi}\left[1-e^{-\phi}\right]^{-1}\exp\left[n_{0}(r+i\psi)\right]\widehat{\Lambda}_{th}(\bar{n})d\psi\,\,.\end{equation}
Taking matrix elements in a number-state basis gives:

\begin{eqnarray}
\left\langle n\right|\widehat{\rho}\left|n'\right\rangle  & = & \frac{\delta_{nn'}}{2\pi}\int_{0}^{2\pi}e^{-(r+i\psi)(n-n_{0})}d\psi\nonumber \\
 & = & \delta_{nn'}\delta_{nn_{0}}\,\,.\label{eq:ftnumberstae}\end{eqnarray}

This effectively Fourier transforms the thermal operator on a circle
of radius $|n_{0}|$ around the origin, thereby generating a pure
number state with boson number equal to $n_{0}$. Thus, extended thermal
bases of this type are certainly able to represent non-Gaussian states
like pure number states. Nevertheless, they cannot represent coherences
between states of different total boson number.

\subsubsection{Thermal operator identities}

The operator identities for the thermal operators are a subset of
the ones obtained previously. There are no useful identities that
map single operators into a differential form, nor are there any for
products like $\widehat{a}_{i}\widehat{a}_{j}$. However, all quadratic
products that involve both annihilation and creation operators have
operator identities.

With this notation, and taking into account the fact that differentiation
with respect to $\mathbf{n}$ now explicitly preserves the skew symmetry
of the generalized variance, the operator identities can be written:

\fbox{\parbox{0.9\columnwidth}{

\begin{eqnarray}
\widehat{\Lambda} & = & \Omega\frac{\partial}{\partial\Omega}\widehat{\Lambda}\nonumber \\
:\widehat{\mathbf{a}}\widehat{\mathbf{a}}^{\dagger}\widehat{\Lambda}: & = & (1+\mathbf{n})\widehat{\Lambda}+(1+\mathbf{n})\frac{\partial\widehat{\Lambda}}{\partial\mathbf{n}}(1+\mathbf{n})\,\,,\nonumber \\
\{\widehat{\mathbf{a}}:\widehat{\mathbf{a}}^{\dagger}\widehat{\Lambda}:\} & = & (1+\mathbf{n})\widehat{\Lambda}+\mathbf{n}\frac{\partial\widehat{\Lambda}}{\partial\mathbf{n}}(1+\mathbf{n})\,\,,\nonumber \\
\{:\widehat{\mathbf{a}}\widehat{\Lambda}:\widehat{\mathbf{a}}^{\dagger}\} & = & (1+\mathbf{n})\widehat{\Lambda}+(1+\mathbf{n})\frac{\partial\widehat{\Lambda}}{\partial\mathbf{n}}\mathbf{n}\,\,,\nonumber \\
\{\widehat{\mathbf{a}}\widehat{\mathbf{a}}^{\dagger}\widehat{\Lambda}\} & = & \mathbf{n}\widehat{\Lambda}+\mathbf{n}\frac{\partial\widehat{\Lambda}}{\partial\mathbf{n}}\mathbf{n}\,\,.\nonumber \\
\label{eq:Thermalidentities}\end{eqnarray}
 }} 

~

\subsection{Coherent projectors}

\subsubsection{Physical states}

Next, we can include coherent displacements of a thermal Gaussian
in the operator basis. This allows us to compare the Gaussian representation
with earlier methods using the simplest type of pure-state basis,
which is the set of coherent states. These have the property that
the variance in position and momentum is fixed, and always set to
the minimal uncertainty values that occur in the ground state of a
harmonic oscillator. 

In general, we consider an $M$-mode bosonic field. In an M-mode bosonic
Hilbert space, the normalized coherent states $\left|\bm\alpha\right\rangle $
are the eigenstates $\left|\bm\alpha\right\rangle $ of annihilation
operators $\widehat{\bm a}$ with eigenvalues $\bm\alpha$. The corresponding
Gaussian density matrices are the coherent pure-state projectors:\begin{equation}
\widehat{\Lambda}_{c}\left(\bm\alpha\right)=\left|\bm\alpha\right\rangle \left\langle \bm\alpha\right|\,\,,\label{eq:GSP-proj}\end{equation}
which are the basis of the Glauber-Sudarshan P-representation. To
compare this with the Gaussian notation, we re-write the projector
using displacement operators, as:\begin{equation}
\widehat{\Lambda}_{c}\left(\bm\alpha\right)=e^{\widehat{\bm a}^{\dagger}\cdot\bm\alpha}\left|\bm0\right\rangle \left\langle \bm0\right|e^{\bm\alpha^{\ast}\cdot\widehat{\bm a}-|\bm\alpha|^{2}}\,\,.\label{eq:Gauss/GS}\end{equation}
Since the vacuum state is an example of a thermal Gaussian, and the
other terms are all normally ordered by construction, this is exactly
the same as the Gaussian operator $\widehat{\Lambda}(1,\bm\alpha,\bm\alpha^{\ast},\bm0,\bm0,\bm0)$
. In other words, if we restrict the Gaussian representation to this
particular subspace, it is identical to the Glauber-Sudarshan P-representation\cite{Gla-P}.
This pioneering technique was very useful in laser physics, as it
directly corresponds to easily measured normally ordered products.
It has the drawback that it is not a complete basis, unless the set
of distributions is allowed to include generalized functions that
are not positive-definite.

Other examples of physical states of this type are the displaced thermal
density operators. These physically correspond to an ideal coherently
generated bosonic mode from a laser or atom laser source, together
with a thermal background. They can be written as:\begin{eqnarray}
\widehat{\Lambda}_{c}\left(\bm\alpha,\overline{\bm n}\right) & = & \widehat{\Lambda}(1,\bm\alpha,\bm\alpha^{\ast},\overline{\bm n},\bm0,\bm0)\nonumber \\
 & = & e^{\widehat{\bm a}^{\dagger}\cdot\bm\alpha-\widehat{\bm a}\bm\alpha^{*}}\widehat{\Lambda}_{\textrm{th}}(\overline{\bm n})e^{\bm\alpha^{\ast}\cdot\widehat{\bm a}-\widehat{\bm a}^{\dagger}\cdot\bm\alpha}\,\,.\label{eq:disptherm}\end{eqnarray}

\subsubsection{Generalized coherent projectors}

There are two ways to generalize the coherent projectors into operators
that are not density matrices: either by altering the thermal boson
number $\overline{\bm n}$ so it does not correspond to a physical
state, or by changing the displacements so they are not complex conjugate
to each other.

The first procedure is the most time-honored one, since it is the
route by which one can generate the classical phase-space representations
that correspond to different operator orderings. The Wigner\cite{Wig-Wigner},
Q-function\cite{Hus-Q}, and s-ordered\cite{CG-Q} bases are very
similar to Gaussian density matrices, except with negative mean boson
numbers:\begin{eqnarray}
\widehat{\Lambda}_{W}\left(\bm\alpha\right) & = & \widehat{\Lambda}(1,\bm\alpha,\bm\alpha^{\ast},-\bm I/2,\bm0,\bm0)\nonumber \\
\widehat{\Lambda}_{Q}\left(\bm\alpha\right) & = & \widehat{\Lambda}(1,\bm\alpha,\bm\alpha^{\ast},-\bm I,\bm0,\bm0)\nonumber \\
\widehat{\Lambda}_{s}\left(\bm\alpha\right) & = & \widehat{\Lambda}(1,\bm\alpha,\bm\alpha^{\ast},\bm I(s-1)/2,\bm0,\bm0)\,\,.\label{eq:classicalphase}\end{eqnarray}
As pointed out in the previous subsection, it is also possible to
choose $\overline{\bm n}$ to be non-Hermitian, which would allow
one to obtain representations of coherently displaced number states.
However, there is a problem with this class of non-normally ordered
representations. Generically, they have a restricted set of operator
identities available, and typically lead to Fokker-Planck equations
of higher than second order - with no stochastic equivalents - when
employed to treat nonlinear Liouville equations.

Another widely used complete basis is the scaled coherent-state projection
operator used in the positive-P representation\cite{DG-PosP} and
its stochastic gauge extensions\cite{gauge_paper}:\begin{equation}
\widehat{\Lambda}_{P}\left(\Omega,\bm\alpha,\bm\beta\right)=\Omega\frac{\left|\bm\alpha\right\rangle \left\langle \bm\beta^{*}\right|}{\left\langle \bm\beta^{*}\right.\left|\bm\alpha\right\rangle }\,\,.\label{eq:+P-proj}\end{equation}
 Here we have introduced $\bm\beta^{\ast}$ as a vector amplitude
for the coherent state $\left|\bm\beta^{*}\right\rangle $, in a similar
notation to that used previously.

This expansion has a complex amplitude $\Omega$, and a dynamical
phase space which is of twice the usual classical dimension. The extra
dimensions are necessary if we wish to include superpositions of coherent
states, which give rise to off-diagonal matrix elements in a coherent
state expansion. To compare this with the Gaussian notation, the projector
is re-written using displacement operators, as:\begin{eqnarray}
\widehat{\Lambda}_{P}\left(\Omega,\bm\alpha,\bm\beta\right) & = & \Omega e^{\widehat{\bm a}^{\dagger}\cdot\bm\alpha}\left|\bm0\right\rangle \left\langle \bm0\right|e^{\bm\beta\cdot\widehat{\bm a}-\bm\beta\cdot\bm\alpha}\,\nonumber \\
 & = & \widehat{\Lambda}(\Omega,\bm\alpha,\bm\beta,\bm0,\bm0,\bm0)\,.\label{eq:lambda-P}\end{eqnarray}

This follows since the vacuum state is an example of a thermal Gaussian,
and the other terms are all normally ordered by construction. From
earlier work\cite{DG-PosP}, it is known that any Hermitian density
matrix $\widehat{\rho}$ can be expanded with positive probability
in the over-complete basis $\widehat{\Lambda}_{P}$, and it follows
that the same is true for $\widehat{\Lambda}(\overrightarrow{\lambda})$. 

The effects of the annihilation and creation operators on the projectors
are obtained using the results for the actions of operators on the
coherent states, giving: 

\fbox{\parbox{0.9\columnwidth}{

\begin{eqnarray}
\widehat{\Lambda} & = & \Omega\frac{\partial}{\partial\Omega}\widehat{\Lambda}\nonumber \\
\widehat{\mathbf{a}}\widehat{\Lambda} & = & \bm\alpha\widehat{\Lambda}\nonumber \\
\widehat{\mathbf{a}}^{\dagger}\widehat{\Lambda} & = & \left[\bm\beta+\frac{\partial}{\partial\bm\alpha}\right]\widehat{\Lambda}\nonumber \\
\widehat{\Lambda}\widehat{\mathbf{a}} & = & \left[\bm\alpha+\frac{\partial}{\partial\bm\beta}\right]\widehat{\Lambda}\nonumber \\
\widehat{\Lambda}\widehat{\mathbf{a}}^{\dagger} & = & \bm\beta\widehat{\Lambda}\,\,.\label{eq:Coherentidentities}\end{eqnarray}
 }} 

~ Note that here one has $\underline{\underline{\sigma}}^{N}=0$,
and thus all the antinormally ordered identities have just coherent
amplitudes without derivatives, in agreement with the general identities
obtained in the previous section. In treating nonlinear time-evolution,
this has the advantage that some fourth-order nonlinear Hamiltonian
evolution can be treated with only second-order derivatives, which
means that stochastic equations can be used. In a similar way, one
can treat some (but not all) quadratic Hamiltonians using deterministic
evolution only. The fact that all derivatives are analytic - which
is possible since $\bm\beta\neq\bm\alpha^{\ast}$ - is an essential
feature in obtaining stochastic equations for these general cases\cite{DG-PosP}.

\subsection{Squeezing projectors }

\subsubsection{Physical states}

The zero-temperature subset of the Gaussian density operators describe
the set of minimum uncertainty states, which in quantum optics are
the familiar squeezed states\cite{Yue76,CavSch85}. These are most
commonly defined as the result of a squeezing operator on a vacuum
state, followed by a coherent displacement:

\begin{equation}
\widehat{\Lambda}_{\textrm{sq}}(\bm\alpha,\bm\xi,0)=\widehat{{D}}(\bm\alpha)\widehat{S}(\bm\xi)\left|0\right\rangle \left\langle 0\right|\widehat{S}(-\bm\xi)\widehat{{D}}(-\bm\alpha)\,\,.\label{squeezer}\end{equation}
The action of the multi-mode squeezing operator on annihilation and
creation operators is to produce `anti-squeezed operators:\begin{eqnarray}
\widehat{\bm b} & = & \widehat{S}(\bm\xi)\widehat{\bm a}\widehat{S}^{\dagger}(\bm\xi)=\bm\mu\widehat{\bm a}+\bm\nu\widehat{\bm a}^{\dagger T}\nonumber \\
\widehat{\bm b}^{\dagger T} & = & \widehat{S}(\bm\xi)\widehat{\bm a}^{\dagger T}\widehat{S}^{\dagger}(\bm\xi)=\bm\mu^{*}\widehat{\bm a}^{\dagger T}+\bm\nu^{*}\widehat{\bm a}\,\,,\label{eq:yuentrans}\end{eqnarray}
 where the Hermitian matrix $\bm\mu(\bm\xi)$ and the symmetric matrix
$\bm\nu(\bm\xi)$ are defined as multi-mode generalizations of hyperbolic
functions\cite{ZhaFenGil90,LoSol93}:

\begin{eqnarray}
\bm\mu & \equiv & I+\frac{{1}}{2!}\bm\xi\bm\xi^{*}+\frac{{1}}{4!}(\bm\xi\bm\xi^{*})^{2}+\,...\;\equiv\cosh(|\bm\xi|)\nonumber \\
\bm\nu & \equiv & \bm\xi+\frac{{1}}{3!}\bm\xi\bm\xi^{*}\bm\xi+\frac{{1}}{5!}(\bm\xi\bm\xi^{*})^{2}\bm\xi+\,...\;\equiv\frac{{\sinh(|\bm\xi|)}}{|\bm\xi|}\bm\xi.\nonumber \\
\label{sinhandcosh}\end{eqnarray}
Note that $\bm\mu$ and $\bm\nu$ obey the hyperbolic relation $\bm\mu\bm\mu-\bm\nu\bm\nu^{*}=\bm I$
and have the symmetry property $\bm\mu^{-1}\bm\nu=\left(\bm\mu^{-1}\bm\nu\right)^{T}=\bm\nu^{*}\bm\mu^{-1}$.
In the physics of Bose-Einstein condensates, $\widehat{\bm b}$ and
$\widehat{\bm b}^{\dagger}$ are just the Bogoliubov annihilation
and creation operators for quasi-particle excitations. 

The Bogoliubov parameters provide a convenient way of characterizing
the minimum uncertainty Gaussian operators. We therefore need to relate
them to the parameters in the Gaussian covariance matrix. First consider
the antinormal density moment for a squeezed state:\begin{eqnarray}
\left\langle \widehat{\bm a}\widehat{\bm a}^{\dagger}\right\rangle  & = & {\rm {Tr}}\left\{ \widehat{\bm a}\widehat{\bm a}^{\dagger}\widehat{\Lambda}_{\textrm{sq}}(\bm\alpha,\bm\xi,0)\right\} \nonumber \\
 & = & \left\langle 0\right|\widehat{S}(-\bm\xi)\widehat{{D}}(-\bm\alpha)\widehat{\bm a}\widehat{\bm a}^{\dagger}\widehat{{D}}^{\dagger}(-\bm\alpha)\widehat{S}^{\dagger}(-\bm\xi)\left|0\right\rangle \nonumber \\
 & = & \left\langle 0\right|\left(\bm\mu\widehat{\bm a}-\bm\nu\widehat{\bm a}^{\dagger}+\bm\alpha\right)\left(\widehat{\bm a}^{\dagger}\bm\mu-\widehat{\bm a}\bm\nu^{\ast}+\bm\alpha^{\ast}\right)\left|0\right\rangle \nonumber \\
 & = & \bm\alpha\bm\alpha^{\ast}+\bm\mu\bm\mu.\label{eq:antinormdensity}\end{eqnarray}
 Similarly, the anomalous moments are:\begin{eqnarray}
\left\langle \widehat{\bm a}\widehat{\bm a}^{T}\right\rangle  & = & \bm\alpha\bm\alpha^{T}-\bm\mu\bm\nu\nonumber \\
\left\langle \widehat{\bm a}^{\dagger T}\widehat{\bm a}^{\dagger}\right\rangle  & = & \bm\alpha^{\ast T}\bm\alpha^{\ast}-\bm\nu^{\ast}\bm\mu.\label{eq:anomdensity}\end{eqnarray}
Comparing these moments to those of the general Gaussian state {[}Eq.~(\ref{moments}){]},
we see that

\begin{eqnarray}
\bm n & = & \bm\nu\bm\nu^{*}\nonumber \\
\bm m & = & -\bm\mu\bm\nu\nonumber \\
\bm m^{*} & = & -\bm\nu^{*}\bm\mu.\label{eq:mnrelations}\end{eqnarray}

The relationship between the different parameterizations can be written
in a compact form if we make the definitions\begin{eqnarray}
\underline{\underline{\mu}} & = & \left(\begin{array}{cc}
\bm\mu & -\bm\nu\\
-\bm\nu^{*} & \bm\mu^{T}\end{array}\right),\nonumber \\
\underline{\underline{\xi}} & = & \left(\begin{array}{cc}
\bm0 & \bm\xi\\
\bm\xi^{*} & \bm0\end{array}\right),\label{eq:matrixdef}\end{eqnarray}
 in terms of which the relations are\begin{eqnarray}
\underline{\underline{\sigma}} & = & \frac{1}{2}\underline{\underline{\mu}}^{2}+\frac{1}{2}\underline{\underline{I}},\nonumber \\
\underline{\underline{\mu}} & = & \exp\left(-\underline{\underline{\xi}}\right).\label{eq:mnrelationsII}\end{eqnarray}
One implication of this relation is that, just as $\bm\mu$ is not
independent of $\bm\nu$, so too $\bm n$ is not independent of $\bm m$
for the squeezed state. From the hyperbolic relation, we see that
$\bm n^{T}=\bm m^{\ast}\left(1+\bm n\right)^{-1}\bm m$. The determinant
of the covariance matrix, required for correct normalization, reduces
to the simpler form:

\begin{eqnarray}
\left|\underline{\underline{\sigma}}\right| & =\left|1+\bm n\right|=\left|\bm\mu\right|^{2} & .\label{eq:covdet}\end{eqnarray}

This set of diagonal squeezing projectors forms the basis that has
previously been used to define squeezed-state based representations\cite{SchCav84}.
Because the basis elements in such bases are not analytic and the
resultant distribution not always positive, these previous representations
suffer from the same deficiency as the Glauber-Sudarshan $P$ representation,
(as opposed to the positive-$P$ representation), i.e. the evolving
quantum state can not always be sampled by stochastic methods.

\subsubsection{Generalized squeezing operators}

A non-Hermitian extension of the squeezed-state basis {[}Eq.~(\ref{squeezer}){]}
can be formed by analytic continuation of its parameters, i.e. by
a replacement of the complex conjugates of $\bm\alpha$ and $\bm\xi$
by independent matrices: $\bm\alpha^{*}\rightarrow\bm\alpha^{+}$
and $\bm\xi^{*}\rightarrow\bm\xi^{+}$. In the Bogoliubov parametrization,
this is equivalent to the replacement $\bm\nu^{*}\rightarrow\bm\nu^{+}$and
to $\bm\mu$ being no longer Hermitian. These non-Hermitian operators
are in the form of off-diagonal squeezing projectors and constitute
the basis of a positive-definite squeezed-state representation. They
include as a special case ($\bm\nu=\bm\nu^{+}=\bm0$, $\bm\mu=\bm I$)
the kernel of the coherent-state positive-$P$ expansion. Thus the
completeness of the more general representation is guaranteed by the
completeness of the coherent-state subset, and we can always find
a positive-$P$ function for any density operator by using the coherent-state
based representation.

\subsection{Thermal squeezing operators }

Mixed (or classical) squeezed states are generated by applying the
squeezing operators to the thermal kernel, rather than to the vacuum
projector:\begin{equation}
\widehat{\Lambda}_{\textrm{sq}}(0,\bm\xi,\overline{\bm n})=\widehat{S}(\bm\xi)\widehat{\Lambda}_{\textrm{th}}(\overline{\bm n})\widehat{\bm\xi S}(-\bm\xi)\,\,.\label{eq:thermalsqueeze}\end{equation}
 In this way, a pure or mixed Gaussian state of arbitrary spread can
be generated. 

Once again, we can relate the covariance parameters characterizing
the final state to the thermal and squeezing parameters by comparing
the moments: \begin{eqnarray}
\left\langle \widehat{\bm a}\widehat{\bm a}^{\dagger}\right\rangle  & = & {\rm {Tr}}\left\{ \widehat{\bm a}\widehat{\bm a}^{\dagger}\widehat{\Lambda}_{{\rm sq}}(0,\bm\xi,\overline{\bm n})\right\} \nonumber \\
 & = & {\rm {Tr}}\left\{ \left(\bm\mu\widehat{\bm a}-\bm\nu\widehat{\bm a}^{\dagger}\right)\left(\widehat{\bm a}^{\dagger}\bm\mu-\widehat{\bm a}\bm\nu^{*}\right)\widehat{\Lambda}_{\textrm{th}}(\overline{\bm n})\right\} \nonumber \\
 & = & \bm\mu\left(\overline{\bm n}+\bm I\right)\bm\mu+\bm\nu\overline{\bm n}^{T}\bm\nu^{*},\label{eq:thermmoms}\end{eqnarray}
since there are no anomalous fluctuations in a thermal state. Similarly,
the squeezing moments are \begin{eqnarray}
\left\langle \widehat{\bm a}\widehat{\bm a}^{T}\right\rangle  & = & -\bm\mu\left(\overline{\bm n}+\bm I\right)\bm\nu-\bm\nu\overline{\bm n}^{T}\bm\mu^{*}\nonumber \\
\left\langle \widehat{\bm a}^{\dagger T}\widehat{\bm a}^{\dagger}\right\rangle  & = & -\bm\mu^{*}\overline{\bm n}^{T}\bm\nu^{*}-\bm\nu^{*}\left(\overline{\bm n}+\bm I\right)\bm\mu.\label{eq:thermmomsII}\end{eqnarray}
 Thus the two parameterizations are related by

\begin{eqnarray}
\bm n & = & \bm\mu\overline{\bm n}\bm\mu+\bm\nu\left(\overline{\bm n}^{T}+\bm I\right)\bm\nu^{*}\nonumber \\
\bm m & = & -\bm\mu\left(\overline{\bm n}+\bm I\right)\bm\nu-\bm\nu\overline{\bm n}^{T}\bm\mu^{*}\nonumber \\
\bm m^{*} & = & -\bm\mu^{*}\overline{\bm n}^{T}\bm\nu^{*}-\bm\nu^{*}\left(\overline{\bm n}+\bm I\right)\bm\mu,\label{eq:mnrelationsIII}\end{eqnarray}
 which can be written in a compact form as\begin{eqnarray}
\underline{\underline{\sigma}} & = & \underline{\underline{\mu}}\left(\underline{\underline{\overline{n}}}+\frac{{1}}{2}\underline{\underline{I}}\right)\underline{\underline{\mu}}+\frac{{1}}{2}\underline{\underline{I}},\label{eq:compactform}\end{eqnarray}
where the thermal matrix is defined as\begin{eqnarray}
\underline{\underline{\overline{n}}} & = & \left(\begin{array}{cc}
\overline{\bm n} & \bm0\\
\bm0 & \overline{\bm n}^{T}\end{array}\right).\label{eq:therm_matrix}\end{eqnarray}

As in the cases for the other bases, these squeezed thermal states
can be analytically continued to form a non-Hermitian basis for a
positive-definite representation. Such a representation would be suited
to Bose-condensed systems, which have a finite-temperature (thermal)
character as well as a quantum (squeezed, or Bogoliubov) character.
Furthermore, the lack of a coherent displacement is natural in atomic
systems, where superpositions of total number are unphysical.

\subsection{Displaced thermal squeezing operators }

Finally, the most general Gaussian density matrix is obtained as stated
earlier, by coherent displacement of a squeezed thermal state:\begin{equation}
\widehat{\Lambda}_{{\rm sq}}(\bm\alpha,\bm\xi,\overline{\bm n})=\widehat{{D}}(\bm\alpha)\widehat{S}(\bm\xi)\widehat{\Lambda}_{{\rm th}}(\overline{\bm n})\widehat{\bm\xi S}(-\bm\xi)\widehat{{D}}(-\bm\alpha)\,\,.\label{eq:dispthermalsqueeze}\end{equation}
 In this way, a pure or mixed Gaussian state of arbitrary location
as well as spread can be generated. In terms of the normally ordered
Gaussian notation, the displacement and covariance of this case are
given by:\begin{eqnarray}
\underline{\underline{\sigma}} & = & \underline{\underline{\mu}}\left(\underline{\underline{\overline{n}}}+\frac{{1}}{2}\underline{\underline{I}}\right)\underline{\underline{\mu}}+\frac{{1}}{2}\underline{\underline{I}},\nonumber \\
\underline{\alpha} & = & \left(\begin{array}{c}
\bm\alpha\\
\bm\alpha^{\ast}\end{array}\right)\,\,.\label{eq:generalgaussparam}\end{eqnarray}

\section{Time evolution}

\label{sec:Time-evolution}The utility of the Gaussian representation
arises when it is used to calculate real or imaginary time evolution
of the density matrix. To understand why it is useful to treat both
types of evolution with the same representation, we recall that the
quantum theory of experimental observations generally requires three
phases: state preparation, dynamical evolution, and measurement. It
is clearly advantageous to carry out all three parts of the calculation
in the same representation, in order that the computed trajectories
and probabilities are compatible throughout. Many-body state preparation
is nontrivial, and often involves coupling to a reservoir, which may
result in a canonical ensemble. This can be computed using imaginary
time evolution, as explained below. Dynamical evolution typically
requires a real-time master equation, while the results of a measurement
process are operator expectation values, which were treated in Section
(III).

\subsection{Operator Liouville Equations}

Either real or imaginary time evolution occurs via a Liouville equation
of generic form:

\begin{equation}
\frac{\partial}{\partial t}\widehat{\rho}(t)=\widehat{L}(\widehat{\rho}(t))\,\,,\end{equation}
 where the Liouville super-operator typically involves pre- and post-multiplication
of $\widehat{\rho}$ by annihilation and creation operators. There
are many examples of this type of equation in physics (and, indeed,
elsewhere). We will consider three generic types of equation here:
imaginary time equations used to construct canonical ensembles, unitary
evolution equations in real time, and general non-unitary equations
used to evolve open systems that are coupled to reservoirs.

We often assume that initially, the steady-state density matrix is
in a canonical or grand canonical ensemble, of form:\begin{equation}
\widehat{\rho}_{u}(\tau)=e^{-\tau\widehat{H}/\hbar}\,\,,\label{eq:canonical}\end{equation}
where $\widehat{\rho}_{u}(\tau)$ is un-normalized, $\tau=\hbar/kT$,
and we can include any chemical potential in the Hamiltonian without
loss of generality. If this is not known exactly, the ensemble can
always be calculated through an evolution equation in $\tau$, whose
intial condition is a known high-temperature ensemble. This equation
can also be expressed as a master equation, though not in Lindblad
form. The resulting equation in `imaginary time', or $\tau$, can
be written using an anti-commutator:\begin{eqnarray}
\hbar\frac{\partial}{\partial\tau}\widehat{\rho}_{u} & = & -\frac{1}{2}\left[\widehat{H},\widehat{\rho}_{u}\right]_{+}.\label{eq:anticomm}\end{eqnarray}
Here the initial condition is just the unit operator.

By comparison, the equation for purely unitary time evolution under
a Hamiltonian $\widehat{H}$ is:\begin{equation}
i\,\hbar\frac{\partial}{\partial t}\widehat{\rho}=\left[\widehat{H},\widehat{\rho}\right]\,\,.\end{equation}

More generally, one can describe either the equilibration of an ensemble,
or non-equilibrium behavior via a master equation representing the
real-time dynamics of a physical system. Equations for damping via
coupling of a system to its environment must satisfy restrictions
to ensure that $\widehat{\rho}$ remains positive-definite. In the
Markovian limit, the resulting form is known as the Lindblad form\cite{GarZol00}:
\begin{eqnarray}
\frac{\partial\widehat{\rho}}{\partial t} & = & -\frac{{i}}{\hbar}\left[\widehat{H},\widehat{\rho}\right]+\sum_{K}\left(2\widehat{O}_{K}\widehat{\rho}\widehat{O}_{K}^{\dagger}-\left[\widehat{\rho},\widehat{O}_{K}^{\dagger}\widehat{O}_{K}\right]_{+}\right),\nonumber \\
\end{eqnarray}
 which consists of a commutator term involving the Hermitian Hamiltonian
operator $\widehat{H}$, as well as damping terms involving an anti-commutator
of the arbitrary operators $\widehat{O}_{k}$.

\subsection{Phase-space mappings}

While the general operator equations become exponentially complex
for large numbers of particles and modes, the use of phase-space mappings
provides a useful tool for mapping these quantum equations of motion
into a form that can be treated numerically, via random sampling techniques.

Using the operator identities in Eq.~(\ref{eq:Matrixidentities}),
one can transform the operator equations in any of these three cases
into an integro-differential equation:\begin{eqnarray}
\frac{\partial\widehat{\rho}(t)}{\partial t} & = & \int P(\overrightarrow{\lambda},t)\left[\mathcal{L}_{A}\widehat{\Lambda}(\overrightarrow{\lambda})\right]d^{p}\overrightarrow{\lambda}\,\,,\label{eq:Integrodifferential}\end{eqnarray}
 where the differential operator $\mathcal{L}_{A}$ is of the general
form 

\begin{equation}
\mathcal{L}_{A}=U+A_{j}\partial_{j}+\frac{1}{2}D_{ij}\partial_{i}\partial_{j}\,,\end{equation}
 with derivative operators to the right, and $i,j=0,...,p$ for the
case of a $p-$parameter Gaussian. We only consider cases where terms
with derivatives of order higher than two do not appear, which implies
a restriction on the nonlinear Hamiltonian structure.

We next apply partial integration to Eq.~(\ref{eq:Integrodifferential}),
which, provided boundary terms vanish, leads to a Fokker-Planck equation
for the distribution:\begin{eqnarray}
\frac{\partial}{\partial t}P(\overrightarrow{\lambda},t) & = & \mathcal{L}_{N}P(\overrightarrow{\lambda},t),\end{eqnarray}
 where the differential operator $\mathcal{L}_{N}$ has derivatives
to the left:\begin{eqnarray}
\mathcal{L}_{N} & = & U-\partial_{j}A_{j}+\frac{1}{2}\partial_{i}\partial_{j}D_{ij}\,.\label{Ln}\end{eqnarray}
Such Fokker-Planck equations have equivalent path-integral and stochastic
forms, which can be treated with random sampling methods.

For example, in the Hamiltonian case, if the original Hamiltonian
$\widehat{H}_{N}(\widehat{\mathbf{a}},\widehat{\mathbf{a}}^{\dagger})$
is normally ordered (annihilation operators to the right), then for
a positive-$P$ representation one can immediately obtain: \begin{equation}
\mathcal{L}_{N}=\frac{1}{i\hbar}\left[H_{N}(\bm\alpha,\bm\beta-\bm\partial_{\alpha})-H_{N}(\bm\beta,\bm\alpha-\bm\partial_{\beta})\right]\,\,.\label{Differential}\end{equation}
With the use of additional identities in $\Omega$ to eliminate the
potential term $U$, the Fokker-Planck equation can be sampled by
stochastic Langevin equations for the phase-space variables. Note
that this potential term only arises with imaginary-time evolution.
The first-order derivative (drift) terms in the Fokker-Planck equation
map to deterministic terms in the Langevin equations, and the second-order
derivative (diffusion) terms map to stochastic terms. To obtain stochastic
equations, we follow the general stochastic gauge technique\cite{gauge_paper},
which in turn is based on the positive-$P$ method. 

To simplify notation, we have left the precise form of the derivatives
in the Fokker-Planck equation as yet unspecified. Different choices
are possible because the Gaussian operator kernel is an analytic function
of its parameters. The standard choice in the positive-$P$ method,
obtained through the dimension-doubling technique\cite{DG-PosP},
is such that when the equation is written in terms of real and imaginary
derivatives, all the coefficients are real and the diffusion is positive-definite.
This ensures that stochastic sampling is always possible. Other choices
are also possible and useful if analytic solutions are desired. 

The structure of the noise terms in the stochastic equations is given
by the noise-matrix $\mathbf{\, B}$, which is defined as a $\, p\times p'$
complex matrix square root: \begin{equation}
\,\mathbf{D}=\mathbf{BB}^{T}\,\,.\end{equation}
Since this is non-unique, one can introduce diffusion gauges from
a set of matrix transformations $\mathbf{U}[\mathbf{f}(\overrightarrow{\lambda})]$
with $\mathbf{UU}^{T}=\mathbf{I}$. It is also possible to introduce
arbitrary drift gauge terms $\,\mathbf{g}$ which are used to stabilize
the resulting stochastic equations:

\begin{eqnarray}
\frac{d\Omega}{dt} & = & \Omega\left[U+\,\mathbf{g}\cdot\bm\zeta(t)\right]\nonumber \\
\frac{d\lambda_{i}}{dt} & = & A_{i}+B_{ij}\left[\zeta_{j}(t)-\, g_{j}\right],\,\,\,\,(i,j>0)\,\,.\label{Central}\end{eqnarray}
These are Ito stochastic equations with noise terms defined by the
correlations:

\begin{equation}
\langle\zeta_{i}(t)\zeta_{j}(t')\rangle=\delta(t-t')\delta_{ij}\,\,.\end{equation}

We note here that the use of stochastic equation sampling as described
here represents only one possible way to sample the underlying Fokker-Planck
equation. Other ways are possible, including the usual Metropolis
and diffusion Monte-Carlo methods found in imaginary-time many-body
theory.

In the remainder of this section, we consider quadratic Hamiltonians,
or master equations. We show that under the Gaussian representation,
these give rise to purely deterministic or `drift' evolution. We first
treat the thermal case, then derive an analytic solution to the dynamics
governed by a general master equation that is quadratic in annihilation
and creation operators. Following this are several examples which
show how the analytic solution can be applied to physical problems.
While these examples can all be treated in other ways, they demonstrate
the technique, which will be extended to higher-order problems subsequently.

\subsection{General quadratic master equations}

Any quadratic master equation can be treated exactly with the Gaussian
distribution. To demonstrate this, we can cast any quadratic master
equation into the form \begin{eqnarray}
\frac{\partial}{\partial t}\widehat{\rho} & = & A^{(0)}\widehat{\rho}+A_{\mu}^{(1)}:\widehat{a}_{\mu}\widehat{\rho}:+B_{\mu}^{(1)}\left\{ \widehat{a}_{\mu}:\widehat{\rho}:\right\} \nonumber \\
 & + & A_{\nu\mu}:\widehat{a}_{\mu}\widehat{a}_{\nu}^{\dagger}\widehat{\rho}:+B_{\nu\mu}\left\{ \widehat{a}_{\mu}\widehat{a}_{\nu}^{\dagger}:\widehat{\rho}:\right\} \nonumber \\
 & + & C_{\nu\mu}\left\{ \widehat{a}_{\mu}:\widehat{a}_{\nu}^{\dagger}\widehat{\rho}:\right\} ,\nonumber \\
 & = & A^{(0)}\widehat{\rho}+\underline{\underline{{\mathrm{Tr}}}}\left[\underline{A}^{(1)}:\widehat{\underline{a}}\widehat{\rho}:+\underline{B}^{(1)}\left\{ \widehat{\underline{a}}:\widehat{\rho}:\right\} \right]\nonumber \\
 & + & \underline{\underline{{\mathrm{Tr}}}}\left[\underline{\underline{A}}:\widehat{\underline{a}}\,\widehat{\underline{a}}^{\dagger}\widehat{\rho}:+\underline{\underline{B}}\left\{ \widehat{\underline{a}}\,\widehat{\underline{a}}^{\dagger}:\widehat{\rho}:\right\} +\underline{\underline{C}}\left\{ \widehat{\underline{a}}\,:\widehat{\underline{a}}^{\dagger}\widehat{\rho}:\right\} \right]\nonumber \\
\label{generalm.e.}\end{eqnarray}
where the trace is a matrix structural operation (indicated by the
double underline), not a trace over the operators. Here $A^{(0)}$
is a real number, while $\underline{A}^{(1)}$and $\underline{B}^{(1)}$
are complex column vectors with the generalised Hermitian property
of $\underline{A}^{(1)*T}=\underline{A}^{(1)+}$, $\underline{B}^{(1)*T}=\underline{B}^{(1)+}$. 

The quadratic terms $\underline{\underline{A}}$, $\underline{\underline{B}}$
and $\underline{\underline{C}}$ are complex-number matrices that
have the implicit superscript $(2)$ dropped for notational simplicity.
By construction, $\underline{\underline{A}}$ and $\underline{\underline{B}}$
possess all the skew symmetries of $\underline{\underline{\sigma}}$:
$\underline{\underline{A}}=\underline{\underline{A}}^{+}$ and $\underline{\underline{B}}=\underline{\underline{B}}^{+}$
, i.e. they are Hermitian in the generalized sense defined earlier.
The matrix $\underline{\underline{C}}$ possesses only some of these
skew symmetries, namely that the upper right and lower left blocks
are each symmetric. Furthermore, the Hermiticity of the density operator
requires that the matrix Hermitian conjugate is equal to the generalized
Hermitian conjugate: $\underline{\underline{A}}^{*T}=\underline{\underline{A}}^{+}$,
$\underline{\underline{B}}^{*T}=\underline{\underline{B}}^{+}$ and
$\underline{\underline{C}}^{*T}=\underline{\underline{C}}^{+}$. 

By expanding $\widehat{\rho}$ in the general Gaussian basis and applying
the operator identities in Eq.~(\ref{eq:Matrixidentities}), we obtain
a Liouville equation for the phase-space distribution $P$ that contains
only zeroth- and first-order derivatives. Since this can be treated
by the method of characteristics, the time-evolution is deterministic:
every initial value corresponds uniquely to a final value, without
diffusion or stochastic behavior. This can also be solved analytically,
since the time-evolution resulting from a quadratic master equation
is linear in the Gaussian parameters $\overrightarrow{\lambda}$.

\subsubsection{Imaginary time evolution}

We consider this case in detail, even though it is relatively straightforward,
because it gives an example of phase-space evolution which would require
diffusive or stochastic equations using previous methods. The equation
in `imaginary time', or $\tau$ can be written using an anti-commutator.
Since we are only considering linear evolution here, the relevant
Hamiltonian is always diagonalizable, and can be written as:

\begin{equation}
\widehat{H}=\hbar:\widehat{\mathbf{a}}^{\dagger}\bm\omega\,\widehat{\mathbf{a}}:\,\,.\end{equation}

Next, we need to cast the un-normalized density operator equation:\begin{eqnarray}
\hbar\frac{\partial}{\partial\tau}\widehat{\rho}_{u} & = & -\frac{1}{2}\left[\widehat{H},\widehat{\rho}_{u}\right]_{+}\end{eqnarray}
into differential form. All the terms are of mixed form, including
both normal and antinormal ordered parts, so the master equation can
be written as:\begin{equation}
\frac{\partial}{\partial\tau}\widehat{\rho}_{u}=\underline{\underline{{\mathrm{Tr}}}}\left[\underline{\underline{C}}\left\{ \widehat{\underline{a}}\,:\widehat{\underline{a}}^{\dagger}\widehat{\rho}_{u}:\right\} \right]+A^{(0)}\widehat{\rho}_{u}\,\,,\end{equation}
where: \begin{eqnarray}
\underline{\underline{C}} & =-\frac{1}{2} & \left[\begin{array}{cc}
\bm\omega & \bm0\\
\bm0 & \bm\omega^{T}\end{array}\right]\nonumber \\
A^{(0)} & = & -{\mathrm{Tr}}\,\bm\omega\,\,.\end{eqnarray}

Using the identities in Eq (\ref{eq:Matrixidentities}), one finds
that the corresponding differential operator to be\begin{equation}
\mathcal{L}_{A}\widehat{\Lambda}=\left[A^{(0)}\widehat{\Lambda}+{\underline{\underline{\mathrm{Tr}}}}\,\underline{\underline{C}}\left(1+\,\underline{\underline{\sigma}}^{N}\frac{\partial}{\partial\underline{\underline{\sigma}}}\,\right)\widehat{\Lambda}\underline{\underline{\sigma}}\right]\,.\end{equation}
This leads to the following equation for the distribution, after integration
by parts (which requires mild restrictions on the initial distribution):
\begin{eqnarray}
\frac{\partial P}{\partial\tau} & = & \sum_{k}\omega_{k}\left[\frac{\partial}{\partial\Omega}\Omega+\frac{\partial}{\partial n_{k}}n_{k}\right]\left(1+n_{k}\right)P\,\,.\end{eqnarray}
Solving first-order Fokker-Planck like equations in this form leads
to deterministic characteristic equations:

\begin{eqnarray}
\dot{\Omega} & = & -\sum_{k}\omega_{k}\Omega\left(1+n_{k}\right)\,\,,\nonumber \\
\dot{n}_{k} & = & -\omega_{k}n_{k}\left(1+n_{k}\right)\,\,.\end{eqnarray}

Integrating the deterministic equation for the mode occupation $n_{k}$
leads to the Bose-Einstein distribution also encountered in Eq (\ref{eq:BoseEinstein}):

\begin{equation}
n_{k}=\frac{1}{e^{\omega_{k}\tau}-1}\,\,.\end{equation}
The weighting term occurs because this method of obtaining a thermal
density matrix results in an un-normalized density matrix with trace
equal to $\Omega(\tau)$. From integration of the above equation one
finds, as expected from Eq (\ref{Th_{u}nordered}), that:

\begin{equation}
Tr\left[\widehat{\rho}_{u}\right]=\Omega(\tau)=\Omega_{0}\Pi_{k}\left[1-e^{-\omega_{k}\tau}\right]^{-1}\,\,.\end{equation}

\subsubsection{Real time evolution}

In the Lindblad form of a master equation which is relevant to real
time evolution, further restrictions apply to its structure than just
the symmetries given above. 

The preservation of the trace of $\widehat{\rho}$ in real-time master
equations requires that $\underline{A}^{(1)}=-\underline{B}^{(1)}$.
In addition, we require that $\underline{\underline{{\mathrm{Tr}}}}\,\underline{\underline{B}}=-\underline{\underline{{\mathrm{Tr}}}}\,(\underline{\underline{A}}+\underline{\underline{C}})=A^{(0)}$
and that the matrix sum $\underline{\underline{D}}=\underline{\underline{A}}+\underline{\underline{B}}+\underline{\underline{C}}$
is anti-skew-symmetric : $\underline{\underline{D}}^{+}=-\underline{\underline{D}}$.
The resultant differential equation for $P$ is simplified by the
fact that most of the symmetric terms from the identities are multiplied
by the antisymmetric $\underline{\underline{D}}$, and thus give a
trace of zero. In particular, the zeroth-order terms will sum to zero,
leaving \begin{eqnarray}
\frac{\partial P}{\partial t} & = & -\left[\frac{\partial}{\partial\underline{\alpha}}\left(\underline{\underline{E}}\,\underline{\alpha}+\underline{A}^{(1)}\right)\right.\nonumber \\
 & + & \left.\underline{\underline{{\mathrm{Tr}}}}\frac{\partial}{\partial\underline{\underline{\sigma}}}\left(2\underline{\underline{B}}+\underline{\underline{E}}\,\underline{\underline{\sigma}}+\underline{\underline{\sigma}}\,\underline{\underline{E}}^{+}\right)\right]P,\nonumber \\
\label{eq:Realtime}\end{eqnarray}
where $\underline{\underline{E}}=2\underline{\underline{A}}+\underline{\underline{C}}=-2\underline{\underline{B}}-\underline{\underline{C}}^{+}$
. 

A Fokker-Planck equation such as this that contains only first-order
derivatives describes a drift of the distribution and can be converted
into equivalent deterministic equations for the phase-space variables:\begin{eqnarray}
\dot{\underline{\alpha}} & = & \underline{A}^{(1)}+\underline{\underline{E}}\underline{\alpha},\nonumber \\
\dot{\underline{\underline{\sigma}}} & = & 2\underline{\underline{B}}+\underline{\underline{E}}\,\underline{\underline{\sigma}}+\underline{\underline{\sigma}}\,\underline{\underline{E}}^{+}.\end{eqnarray}
 This system of linear ordinary differential equations has the general
solution\begin{eqnarray}
\underline{\alpha}(t) & = & e^{\underline{\underline{E}}t}\left(\underline{\alpha}(0)-\underline{\alpha}^{0}\right)+\underline{\alpha}^{0}\nonumber \\
\underline{\underline{\sigma}}(t) & = & e^{\underline{\underline{E}}t}\left(\underline{\underline{\sigma}}(0)-\underline{\underline{\sigma}}^{0}\right)e^{\underline{\underline{E}}^{+}t}+\underline{\underline{\sigma}}^{0},\label{generalsolution}\end{eqnarray}
 where $\underline{\alpha}^{0}$ satisfies $\underline{\underline{E}}\,\underline{\alpha}^{0}=-\underline{A}^{(1)}$,
and the skew-symmetric matrix $\underline{\underline{\sigma}}^{0}$
satisfies $\underline{\underline{E}}\,\underline{\underline{\sigma}}^{0}+\underline{\underline{\sigma}}^{0}\,\underline{\underline{E}}^{+}=-2\underline{\underline{B}}$.
Note that if $\underline{\underline{E}}$ is Hermitian and negative
definite, then the dynamics will consist of some initial transients
with a decay to the steady state: $\underline{\alpha}(\infty)=\underline{\alpha}^{0}$,
$\underline{\underline{\sigma}}(\infty)=\underline{\underline{\sigma}}^{0}$. 

The first- and second-order physical moments also have a simple analytic
form:\begin{eqnarray}
\left\langle \underline{\widehat{a}}\right\rangle (t) & = & e^{\underline{\underline{E}}t}\left(\left\langle \underline{\widehat{a}}\right\rangle (0)-\underline{\alpha}^{0}\right)+\underline{\alpha}^{0}\nonumber \\
\left\langle :\underline{\widehat{a}}\,\underline{\widehat{a}}^{\dagger}:\right\rangle (t) & = & e^{\underline{\underline{E}}t}\left(\left\langle :\underline{\widehat{a}}\,\underline{\widehat{a}}^{\dagger}:\right\rangle (0)-\underline{\underline{F}}(0)\right)e^{\underline{\underline{E}}^{+}t}+\underline{\underline{F}}(t)\,,\nonumber \\
\end{eqnarray}
where the steady state with coherent transients is given by: \begin{equation}
\underline{\underline{F}}(t)=\underline{\underline{\sigma}}^{0}+\left\langle \underline{\widehat{a}}^{\,}\right\rangle (t)\,\left\langle \underline{\widehat{a}}^{\dagger}\right\rangle (t)-\underline{\underline{I}}\,\,.\end{equation}

For a quadratic master equation in Lindblad form, the Hamiltonian
and damping operators can be expressed as\begin{eqnarray}
\widehat{H} & = & \underline{\underline{{\mathrm{Tr}}}}\,\left(\underline{\underline{H}}:\underline{\widehat{a}}\,\underline{\widehat{a}}^{\dagger}:\right)\nonumber \\
\widehat{O}_{K} & = & \underline{O}_{K}^{*}\,\underline{\widehat{a}}\nonumber \\
\widehat{O}_{K}^{\dagger} & = & \underline{\widehat{a}}^{\dagger}\,\underline{O}_{K},\label{eq:Lindbladops}\end{eqnarray}
 where, in block form, \begin{eqnarray}
\underline{\underline{H}} & = & \left(\begin{array}{cc}
\bm H^{(1)} & \bm H^{(2)}\\
\bm H^{(2)*} & \bm H^{(1)T}\end{array}\right)\nonumber \\
\underline{O}_{K} & = & \left(\begin{array}{c}
\bm O_{K}^{(1)}\\
\bm O_{K}^{(2)}\end{array}\right)\nonumber \\
\underline{O}_{K}^{*} & = & \left(\begin{array}{cc}
\bm O_{K}^{(1)*} & \bm O_{K}^{(2)*}\end{array}\right).\label{eq:Lindbladmatrices}\end{eqnarray}
 Thus the coefficients of density terms $:\widehat{\bm a}\widehat{\bm a}^{\dagger}:$
appear in $\bm H^{(1)}$ and the coefficients of squeezing terms $\widehat{\bm a}\widehat{\bm a}^{T}$
appear in $\bm H^{(2)}$. The commutator term and each of the damping
terms will provide a contribution to the matrices $\underline{\underline{A}}$,
$\underline{\underline{B}}$, and $\underline{\underline{C}}$, which
we can label respectively as $\underline{\underline{A}}_{H}$, $\underline{\underline{A}}_{K}$,
etc. The contributions from the Hamiltonian term are thus:\begin{eqnarray}
\underline{\underline{A}}_{H} & = & \left(\begin{array}{cc}
\bm0 & i\bm H^{(2)}\\
-i\bm H^{(2)*} & \bm0\end{array}\right),\nonumber \\
\underline{\underline{C}}_{H} & = & \left(\begin{array}{cc}
-2i\bm H^{(1)} & \bm0\\
\bm0 & 2i\bm H^{(1)T}\end{array}\right),\label{eq:ACexp}\end{eqnarray}
 and $\underline{\underline{B}}_{H}=-\underline{\underline{A}}_{H}$.
With only the Hamiltonian (unitary) contributions, the matrix $\underline{\underline{E}}$
appearing in the general solution is anti-Hermitian. 

In contrast, the contributions from the damping terms are \begin{eqnarray}
\underline{\underline{A}}_{K} & = & \left(\begin{array}{cc}
\left(\bm O_{K}^{(2)}\bm O_{K}^{(2)*}\right)^{T} & -\bm O_{K}^{(1)}\bm O_{K}^{(2)*}\\
-\bm O_{K}^{(2)}\bm O_{K}^{(1)*} & \bm O_{K}^{(2)}\bm O_{K}^{(2)*}\end{array}\right),\nonumber \\
\underline{\underline{B}}_{K} & = & \left(\begin{array}{cc}
\bm O_{K}^{(1)}\bm O_{K}^{(1)*} & -\bm O_{K}^{(1)}\bm O_{K}^{(2)*}\\
-\bm O_{K}^{(2)}\bm O_{K}^{(1)*} & \left(\bm O_{K}^{(1)}\bm O_{K}^{(1)*}\right)^{T}\end{array}\right),\label{eq:ABexp}\end{eqnarray}
and $\underline{\underline{C}}_{K}=-\underline{\underline{A}}_{K}-\underline{\underline{B}}_{K}$.
With only these damping contributions, the matrix $\underline{\underline{E}}$
is Hermitian.

\subsection{Bogoliubov dynamics}

As an example of how the dynamics of a linear problem can be solved
exactly with these methods, consider the following quadratic Hamiltonian\begin{equation}
\widehat{H}_{s}=\hbar\sum_{i,j=1}^{m}\frac{{i}}{2}\left[\chi_{ij}\widehat{a}_{i}^{\dagger}\widehat{a}_{j}^{\dagger}-\chi_{ij}^{*}\widehat{a}_{i}\widehat{a}_{j}\right],\label{Bogoliubov}\end{equation}
 where $\bm\chi$ is a complex symmetric matrix. In the single-mode
case, this Hamiltonian describes two-photon down-conversion from an
undepleted (classical) pump\cite{WalMil94}. The full multi-mode model
describes quasi-particle excitation in a BEC within the Bogoliubov
approximation\cite{leg01}. Alternatively, it may be used to describe
the dissociation of a large molecular condensate into its constituent
atoms\cite{Kherunts}. Recasting this system into the general master-equation
form {[}Eq.~(\ref{generalm.e.}){]}, we find that the constant and
linear terms vanish, $\underline{\underline{C}}=0$ and\begin{equation}
\underline{\underline{E}}=-2\underline{\underline{B}}=2\underline{\underline{A}}=\left[\begin{array}{cc}
\bm0 & \bm\chi^{*}\\
\bm\chi & \bm0\end{array}\right].\end{equation}
 The general solution {[}Eq.~(\ref{generalsolution}){]} can then
be written\begin{eqnarray}
\underline{\alpha}(t) & = & \left[\begin{array}{cc}
\cosh^{*}\left(|\bm\chi|t\right) & \bm\chi^{*}\frac{\sinh\left(|\bm\chi|t\right)}{|\bm\chi|}\\
\frac{\sinh\left(|\bm\chi|t\right)}{|\bm\chi|}\bm\chi & \cosh\left(|\bm\chi|t\right)\end{array}\right]\underline{\alpha}(0),\nonumber \\
\underline{\underline{\sigma}}(t) & = & \left[\begin{array}{cc}
\cosh^{*}\left(|\bm\chi|t\right) & \bm\chi^{*}\frac{\sinh\left(|\bm\chi|t\right)}{|\bm\chi|}\\
\frac{\sinh\left(|\bm\chi|t\right)}{|\bm\chi|}\bm\chi & \cosh\left(|\bm\chi|t\right)\end{array}\right]\left(\underline{\underline{\sigma}}(0)-\frac{1}{2}\underline{\underline{I}}\right)\nonumber \\
 &  & \left[\begin{array}{cc}
\cosh^{*}\left(|\bm\chi|t\right) & \bm\chi^{*}\frac{\sinh\left(|\bm\chi|t\right)}{|\bm\chi|}\\
\frac{\sinh\left(|\bm\chi|t\right)}{|\bm\chi|}\bm\chi & \cosh\left(|\bm\chi|t\right)\end{array}\right]+\frac{1}{2}\underline{\underline{I}},\end{eqnarray}
where the matrix cosh and sinh functions are as defined in Eq.~(\ref{sinhandcosh}).
If the system starts in the vacuum, for example, then the first-order
moments will remain zero, whereas the second-order moments will grow
as\begin{eqnarray}
\left\langle :\widehat{\bm a}\widehat{\bm a}^{\dagger}:\right\rangle  & = & \frac{1}{2}\cosh^{*}\left(2|\bm\chi|t\right)-\frac{1}{2}\bm I,\nonumber \\
\left\langle \widehat{\bm a}\widehat{\bm a}^{T}\right\rangle  & = & \frac{1}{2}\bm\chi^{*}\frac{\sinh\left(2|\bm\chi|t\right)}{|\bm\chi|}.\end{eqnarray}

\subsection{Dynamics of a Bose gas in a lossy trap}

As a second example, we consider a trapped, noninteracting Bose gas
with loss modeled by an inhomogeneous coupling to a zero-temperature
reservoir\cite{GarZol00}:\begin{equation}
\frac{\partial}{\partial t}\widehat{\rho}=-i\left[\omega_{ij}\widehat{a}_{i}^{\dagger}\widehat{a}_{j},\widehat{\rho}\right]+\frac{1}{2}\gamma_{ij}\left(2\widehat{a_{i}}\widehat{\rho}\widehat{a_{j}}^{\dagger}-\widehat{a_{j}}^{\dagger}\widehat{a_{i}}\widehat{\rho}-\widehat{\rho}\widehat{a_{j}}^{\dagger}\widehat{a_{i}}\right),\end{equation}
where $\bm\omega$ is an Hermitian matrix that describes the mode
couplings and frequencies of the isolated system, and $\bm\gamma$
is an Hermitian matrix that describes the inhomogeneous atom loss.
Recasting this in the general form, we find that $A^{(0)}={\mathrm{Tr}}\,\bm\gamma$,
$\underline{\underline{A}}=0$ and\begin{equation}
\underline{\underline{B}}=\frac{1}{2}\left[\begin{array}{cc}
\bm\gamma^{T} & 0\\
0 & \bm\gamma\end{array}\right],\,\,\,\,\,\underline{\underline{C}}=-i\left[\begin{array}{cc}
\widetilde{\bm\omega} & 0\\
0 & -\widetilde{\bm\omega}^{\ast}\end{array}\right],\end{equation}
 where $\widetilde{\bm\omega}=\bm\omega-i\bm\gamma^{T}/2$. The block-diagonal
form of these matrices allows us to write the solution to the phase-space
equations as\begin{eqnarray}
\bm\alpha(t) & = & e^{-i\widetilde{\bm\omega}t}\bm\alpha(0),\nonumber \\
\bm\alpha^{+}(t) & = & \bm\alpha^{+}(0)e^{i\widetilde{\bm\omega}^{\dagger}t},\nonumber \\
\bm n(t) & = & e^{-i\widetilde{\bm\omega}t}\bm n(0)e^{i\widetilde{\bm\omega}^{\dagger}t}\nonumber \\
\bm m(t) & = & e^{-i\widetilde{\bm\omega}t}\bm m(0)e^{-i\widetilde{\bm\omega}^{T}t}\nonumber \\
\bm m^{+}(t) & = & e^{i\widetilde{\bm\omega}^{\dagger T}t}\bm m^{+}(0)e^{i\widetilde{\bm\omega}^{\dagger}t}.\label{eq:Bosesoln}\end{eqnarray}
 If $\bm\gamma$ is positive definite \textit{}and commutes with $\bm\omega$,
then the dynamics will be transient, and all these moments will decay
to zero.

\subsection{Parametric amplifier}

A single-mode example that includes features of the previous two systems
is a parametric amplifier consisting of a single cavity mode parametrically
pumped (at rate $\chi$) via down-conversion of a classical input
field and subject to one-photon loss (at rate $\gamma$)\cite{KinDru91}:\begin{equation}
\frac{\partial}{\partial t}\widehat{\rho}=\frac{1}{2}\left[\chi\widehat{a}^{\dagger}\widehat{a}^{\dagger}-\chi^{*}\widehat{a}\widehat{a},\rho\right]+\frac{1}{2}\gamma\left(2\widehat{a}\widehat{\rho}\widehat{a}^{\dagger}-\widehat{a}^{\dagger}\widehat{a}\widehat{\rho}-\widehat{\rho}\widehat{a}^{\dagger}\widehat{a}\right).\end{equation}
 This corresponds to phase-space equations with\begin{equation}
\underline{\underline{A}}=\frac{1}{2}\left[\begin{array}{cc}
0 & \chi^{*}\\
\chi & 0\end{array}\right],\,\,\underline{\underline{B}}=\frac{1}{2}\left[\begin{array}{cc}
\gamma & -\chi^{*}\\
-\chi & \gamma\end{array}\right],\,\,\underline{\underline{C}}=-\frac{\gamma}{2}\left[\begin{array}{cc}
1 & 0\\
0 & 1\end{array}\right],\end{equation}
 giving the solutions, for real $\chi$,\begin{eqnarray}
\underline{\alpha}(t) & = & e^{-\gamma\, t/2}\left[\begin{array}{cc}
\cosh\chi t & \sinh\chi t\\
\sinh\chi t & \cosh\chi t\end{array}\right]\underline{\alpha}(0)\nonumber \\
\underline{\underline{\sigma}}(t) & = & e^{-\gamma\, t/2}\left[\begin{array}{cc}
\cosh\chi t & \sinh\chi t\\
\sinh\chi t & \cosh\chi t\end{array}\right]\left(\underline{\underline{\sigma}}(0)-\underline{\underline{\sigma}}^{0}\right)\nonumber \\
 &  & \left[\begin{array}{cc}
\cosh\chi t & \sinh\chi t\\
\sinh\chi t & \cosh\chi t\end{array}\right]e^{-\gamma t/2}+\underline{\underline{\sigma}}^{0}\nonumber \\
\underline{\underline{\sigma}}^{0} & = & \frac{1}{\gamma^{2}-4\chi^{2}}\left[\begin{array}{cc}
\gamma^{2}-2\chi^{2} & \chi\gamma\\
\chi\gamma & \gamma^{2}-2\chi^{2}\end{array}\right],\label{eq:parampsoln}\end{eqnarray}
 which are valid for $\gamma\neq2\chi$. For $\gamma>2\chi$, the
system reaches the steady state $\underline{\alpha}=\underline{0}$,
$\underline{\underline{\sigma}}=\underline{\underline{\sigma}}^{0}$,
i.e. $\left\langle \widehat{a}^{\dagger}\widehat{a}\right\rangle ^{0}=n^{0}=2\chi^{2}/(\gamma^{2}-4\chi^{2})$
and $\left\langle \widehat{a}\widehat{a}\right\rangle ^{0}=m^{0}=\chi\gamma/(\gamma^{2}-4\chi^{2})$.

While this result is well-known and can be obtained in other ways\cite{DruMcNWal81,KinDru91},
it is important to understand the significance of the result in terms
of phase-space distributions. In all previous approaches to this problem
using phase-space techniques, the dynamically changing variances meant
that all distributions would necessarily have a finite width and thus
a finite sampling error. However, the Gaussian phase-space representation
is able to handle all the linear terms in the master equation simply
by adjusting the variance of the basis set. This implies that there
is no sampling error in a numerical simulation of this problem. Sampling
error can only occur if there are nonlinear terms in the master equation.
These issues relating to nonlinear evolution will be treated in a
subsequent publication.

\section{Conclusion}

The operator representations introduced here represent the largest
class of bosonic representations that can be constructed using an
operator basis that is Gaussian in the elementary annihilation and
creation operators. In this sense, they give an appropriate generalization
to the phase-space methods that started with the Wigner representation.
There are a number of advantages inherent in this enlarged class. 

Since the basis set is now very adaptable, it allows a closer match
between the physical density matrix and appropriately chosen members
of the basis. This implies that it should generally be feasible to
have a relatively much narrower distribution over the basis set for
any given density matrix. Thus, there can be great practical advantages
in using this type of basis for computer simulations. Sampling errors
typically scale as $1/\sqrt{T}$ for an ensemble of $T$ trajectories,
so reducing the sampling error gives potentially a quadratic improvement
in the simulation time through reduction in the ensemble size. As
many-body simulations are extremely computer-intensive, both in real
and imaginary time, this could provide substantial improvements. Given
the currently projected limitations on computer hardware performance,
improvement through basis refinement may prove essential in practical
simulations.

We have derived the identities which are essential for first-principles
calculations of the time evolution of quantum systems, both dynamical
(real time) and canonical (imaginary time). Any quadratic master equation
has an exact solution than can be written down immediately from the
general form that we have derived. Higher order problems with nonlinear
time-evolution can be solved by use of stochastic sampling methods,
since we have shown that all Hamiltonians up to quartic order can
be transformed into a second-order Fokker-Planck equation, provided
a suitable gauge is chosen that eliminates all boundary terms. Because
the Gaussian basis is analytic, methods previously used for the stochastic
gauge positive-P representation are therefore applicable for the development
of a positive semi-definite diffusion and corresponding stochastic
equations\cite{DG-PosP,gauge_paper} here. The ability to potentially
transform \emph{all} possible Hamiltonians of quartic order into stochastic
equations did not exist in previous representations.

However, we can point already to a clear advantage to the present
method in terms of deterministic evolution. For example, the initial
condition and complete time evolution of either a squeezed state (linear
evolution in real time) or a thermal state (linear evolution in imaginary
time), with a quadratic master equation, is totally deterministic
with the present method. By comparison, any previously used technique
would result in stochastic equations or stochastic initial conditions,
with a finite sampling error, in either case. While this is not an
issue when treating problems with a known analytic solution, it means
that in more demanding problems it is possible to develop simulation
techniques in which the quadratic terms only give rise to deterministic
rather than random contributions to the simulation, thus removing
the corresponding sampling errors.

Finally, we note that the generality of the Gaussian formalism opens
up the possibility of extending these representations to fermionic
systems.

\begin{acknowledgments}
We gratefully acknowledge useful discussions with K. V. Kheruntsyan
and M. J. Davis. Funding for this research was provided by an Australian
Research Council Centre of Excellence grant.
\end{acknowledgments}
\appendix

\section{Bosonic identities }

\label{sec:Bosonic-identities}To obtain the operator identities required
to treat the time evolution of a general Gaussian operator, we need
a set of theorems and results about operator commutators. These can
then be used to obtain the result of the action of any given quadratic
operator on any Gaussian operator, as described in the main text.
We use the following bosonic identities which are known in the literature,
but reproduced here for ease of reference:

\begin{description}
\item [Commutation]Theorem (I):\\
Given an analytic function $p(\widehat{\mathbf{a}})$ with a power
series expansion valid everywhere, the following commutation rules
hold:\begin{eqnarray}
\left[p(\widehat{\mathbf{a}}),\,\widehat{a_{i}}^{\dagger}\right] & = & \frac{\partial p(\widehat{\mathbf{a}})}{\partial\widehat{a_{i}}}\,\,,\nonumber \\
\left[\widehat{a_{i}},\, p(\widehat{\mathbf{a}}^{\dagger})\right] & = & \frac{\partial p(\widehat{\mathbf{a}}^{\dagger})}{\partial\widehat{a_{i}}^{\dagger}}\,\,.\label{A_{C}OMMUTE}\end{eqnarray}
Proof: Using a Taylor series expansion of $p(\widehat{\mathbf{a}})$
around the origin in $\,\widehat{a_{i}}$, one can evaluate the commutator
of each term in the power series. Hence,\begin{eqnarray}
\left[p(\widehat{\mathbf{a}}),\,\widehat{a_{i}}^{\dagger}\right] & = & \sum_{n}\left[\frac{\widehat{p}_{n}}{n!}\widehat{a}_{i}^{n},\,\widehat{a_{i}}^{\dagger}\right]\nonumber \\
 & = & \sum_{n>0}\frac{\widehat{p}_{n}}{(n-1)!}\widehat{a}_{i}^{n-1}\nonumber \\
 & = & \frac{\partial p(\widehat{\mathbf{a}})}{\partial\widehat{a_{i}}}\,\,.\label{A_{C}OMMUTE_{P}}\end{eqnarray}
\\
The second result follows by taking the Hermitian conjugate.
\item [Ordering]Theorem (II):\\
Given any analytic normally ordered operator function $p(\widehat{\mathbf{a}}^{\dagger},\widehat{\mathbf{a}})$
with a power series expansion, the following ordering rules hold:\begin{eqnarray}
p(\widehat{\mathbf{a}}^{\dagger},\widehat{\mathbf{a}})(\widehat{a}_{i}^{\dagger})^{n} & = & \left[\widehat{a}_{i}^{\dagger}+\frac{\partial}{\partial\widehat{a_{i}}}\right]^{n}p(\widehat{\mathbf{a}}^{\dagger},\widehat{\mathbf{a}})\,\,,\nonumber \\
(\widehat{a}_{i})^{n}p(\widehat{\mathbf{a}}^{\dagger},\widehat{\mathbf{a}}) & = & p(\widehat{\mathbf{a}}^{\dagger},\widehat{\mathbf{a}})\left[\widehat{a}_{i}+\overleftarrow{\frac{\partial}{\partial\widehat{a_{i}}^{\dagger}}}\right]^{n}\,\,.\label{A_ORDER}\end{eqnarray}
\\
Here the left arrow of the differential operator indicates the direction
of differentiation.\\
We can write these two identities in a unified form by introducing
an antinormal ordering bracket, denoted $\{:\widehat{p}:\widehat{a}\}$,
which places all operators in antinormal order relative to the normal
term $:\widehat{p}:$. With this notation, we can write a single ordering
rule for all cases:\begin{equation}
\left\{ :p(\underline{\widehat{a}}):\widehat{a}_{\mu}\right\} =:\left[\widehat{a}_{\mu}+\frac{\partial}{\partial\widehat{a}_{\mu}^{\dagger}}\right]p(\underline{\widehat{a}}):\,\,.\label{A_ORDER_GEN}\end{equation}

\item [Proof:]Since $\widehat{a_{i}}$ commutes with all other annihilation
operators and $\widehat{a_{i}}^{\dagger}$ commutes with creation
operators, Theorem (I) also holds for any normally ordered operator
$p(\widehat{\mathbf{a}}^{\dagger},\widehat{\mathbf{a}})$, with a
power series expansion; provided derivatives are interpreted as normally
ordered also. The first case above then follows directly from Theorem
(I):\begin{equation}
p(\widehat{\mathbf{a}}^{\dagger},\widehat{\mathbf{a}})\widehat{a}_{i}^{\dagger}=\left[\widehat{a}_{i}^{\dagger}+\frac{\partial}{\partial\widehat{a_{i}}}\right]p(\widehat{\mathbf{a}}^{\dagger},\widehat{\mathbf{a}})\,\,.\label{A_{O}RDER_{P}}\end{equation}
The required result then follows by using the equation above $n$
times, recursively\emph{.} The second result is the Hermitian conjugate
of the first. The last result, Eq (\ref{A_ORDER_GEN}) is simply a
unified form that recreates the previous two equations. This can be
applied recursively, since the RHS of this equation is always normally
ordered by construction.
\item [\emph{Corollary}:]The antinormal combination of a Gaussian operator
$\widehat{\Lambda}_{g}(\underline{\widehat{a}})$ and any single creation
or annihilation operator is given by a direct application of the ordering
theorem, Eq (\ref{A_ORDER}):\begin{equation}
\left\{ :\widehat{\Lambda}_{g}(\underline{\widehat{a}}):\widehat{a}_{\mu}\right\} =:\left[\widehat{a}_{\mu}-\sigma_{\mu\nu}^{-1}\delta\widehat{a}_{\nu}\right]\widehat{\Lambda}_{g}(\underline{\widehat{a}}):\,\,.\end{equation}
\\
It should be noticed here that the above expression assumes the covariance
has the usual symmetry: then every operator occurs twice in the Gaussian
quadratic term, which cancels the factor of two in the exponent.
\end{description}
In the main text, these results are used directly to obtain all the
required operator identities on the Gaussian operators.

\section{Gaussian Integrals}

\label{sec:Gaussian-Integrals}In deriving the normalization, moments
and operator identities of the Gaussian representations, we have had
to calculate nonstandard integrals of complex, multidimensional Gaussian
functions. The basic Gaussian integral that must be evaluated is of
the form\begin{equation}
I=\int d^{2M}\mathbf{z}e^{-\delta\underline{z}^{+}\underline{\underline{\sigma}}^{-1}\delta\underline{z}/2},\end{equation}
 where, as in Eq.~(\ref{Trace}), the covariance $\underline{\underline{\sigma}}$
is a $2M\times2M$ non-Hermitian matrix, and $\delta\underline{z}$
and $\delta\underline{z}^{+}$ are complex vectors of length $2M$.
There are two major differences between this expression and the better
known form of the Gaussian integral. First, the vectors $\delta\underline{z}=\underline{z}-\underline{\alpha}$
and $\delta\underline{z}^{+}=\underline{z}^{\ast}-\underline{\alpha}^{+}$
contain offsets which are not complex-conjugate: $\underline{\alpha}^{\ast}\neq\underline{\alpha}^{+}$.
Second, the vector $\underline{z}$ does not consist of $2M$ independent
complex numbers. Rather, it contains $M$ independent complex numbers
$\mathbf{z}$ and their conjugates $\mathbf{z}^{\ast}$. 

To evaluate such an integral, we first write it explicitly in terms
of real variables as\begin{eqnarray}
I & = & \int d^{2M}\underline{z}e^{-\delta\underline{z}^{+}\underline{\underline{\sigma}}^{-1}\delta\underline{z}/2}\nonumber \\
 & = & \int d^{2M}\underline{x}e^{-\left(\underline{x}^{T}-\underline{x}_{0}^{T}\right)\underline{\underline{\tau}}^{-1}\left(\underline{x}-\underline{x}_{0}\right)/2}\,\,,\end{eqnarray}
where $\underline{x}=\underline{\underline{L}}\,\underline{z}=\left({\rm Re}\,\mathbf{z},{\rm Im}\,\mathbf{z}\right)$,
$\underline{x}_{0}=\underline{\underline{L}}\,\underline{\alpha}=\left((\bm\alpha+\bm\alpha^{+})/2,(\bm\alpha-\bm\alpha^{+})/2i\right)$,
and $\underline{\underline{\tau}}=\underline{\underline{L}}\,\underline{\underline{\sigma}}\,\underline{\underline{L}}^{\dagger}$,
with the transformation matrix\begin{equation}
\underline{\underline{L}}=\frac{1}{2}\left(\begin{array}{cc}
\mathbf{I} & \mathbf{I}\\
-i\mathbf{I} & i\mathbf{I}\end{array}\right).\end{equation}
Note that the offset vector $\underline{x}_{0}$ will be complex,
unless $\bm\alpha^{\ast}=\bm\alpha^{+}$. We may remove it by changing
variables $\underline{u}=\underline{x}-\underline{x}_{0}$ and using
contour integration methods to convert the integral back into an integral
on a real manifold. 

With the offset removed, the square of the integral can be written
in the form of a standard multidimensional Gaussian:\begin{eqnarray}
I^{2} & = & \int d^{2M}\underline{x}d^{2M}\underline{y}e^{-\underline{x}^{T}\,\underline{\underline{\tau}}^{-1}\,\underline{x}/2}e^{-\underline{y}^{T}\,\underline{\underline{\tau}}^{-1}\,\underline{y}/2}\nonumber \\
 & = & \int d^{4M}\underline{u}e^{-\underline{u}^{\ast}\,\underline{\underline{\tau}}^{-1}\,\underline{u}/2},\end{eqnarray}
where $\underline{u}=\underline{x}+i\underline{y}$. Assuming that
the matrix $\underline{\underline{\tau}}^{-1}$ can be diagonalized:
$\underline{\underline{\lambda}}=\underline{\underline{U}}\,\underline{\underline{\tau}}^{-1}\,\underline{\underline{U}}^{\dagger}$,
we can factor the integral into a product of $2M$ integrals over
the complex plane:\begin{eqnarray}
I^{2} & = & \prod_{\mu=1}^{2M}\int d^{2}w_{\mu}e^{-w_{\mu}^{\ast}\lambda_{\mu\mu}w_{\mu}/2},\end{eqnarray}
where $\underline{w}=\underline{\underline{U}}\,\underline{u}$. These
integrals can be evaluated by a transformation to radial coordinates,
giving \begin{eqnarray}
I^{2} & = & \prod_{\mu=1}^{2M}\frac{2\pi}{\lambda_{\mu\mu}}\nonumber \\
 & = & \left(2\pi\right)^{2M}\left|\underline{\underline{\tau}}\right|\,\,,\end{eqnarray}
 which holds provided that all the ${\rm Re}\,\lambda_{\mu\mu}\geq0$,
i. e. that all eigenvalues of $\underline{\underline{\tau}}$ have
positive real part. Finally, noting that $\left|\underline{\underline{\tau}}\right|=\left|\underline{\underline{L}}^{-1}\right|\left|\underline{\underline{\sigma}}\right|\left|\underline{\underline{L}}^{-1\dagger}\right|=2^{-2M}\left|\underline{\underline{\sigma}}\right|$,
we find that\begin{eqnarray}
I & = & \pi^{M}\sqrt{\left|\underline{\underline{\sigma}}\right|}\,\,,\end{eqnarray}
with the condition that the eigenvalues of $\left|\underline{\underline{\sigma}}\right|$
have positive real part.

\end{document}